\documentclass[10pt, twocolumn, comsoc]{IEEEtran}

\usepackage{graphicx,epsfig}
\usepackage[noadjust]{cite}
\usepackage{mcite}
\usepackage{amsfonts,helvet}
\usepackage{fancyhdr}
\usepackage{threeparttable}
\usepackage{epsf,epsfig}
\usepackage{amsthm}
\usepackage{amsmath}
\usepackage{siunitx}
\usepackage{amssymb}
\usepackage{dsfont}
\usepackage{subfigure}
\usepackage{color}
\usepackage[linesnumbered,ruled,noend]{algorithm2e}
\usepackage{algpseudocode}
\usepackage{algcompatible}
\usepackage{enumerate}
\usepackage{gensymb}
\usepackage{cancel}
\usepackage{graphicx}
\usepackage{wrapfig}
\usepackage{ragged2e}

\newtheorem{corollary}{Corollary}

\newtheorem{lemma}{Lemma}

\newtheorem{proposition}{Proposition}
\newtheorem{remark}{Remark}

\usepackage{bbm}
\usepackage{eucal}

\usepackage{dsfont}
\usepackage{boldline}

\begin{document}

\title{ 
Joint and Robust Beamforming Framework for Integrated Sensing and Communication Systems
}


\author{Jinseok~Choi, Jeonghun~Park, Namyoon~Lee, and Ahmed Alkhateeb

\thanks{
J. Choi is with the School of Electrical Engineering, Korea Advanced Institute of Science and Technology (KAIST), Daejeon, 34141, South Korea (e-mail: {\texttt{jinseok@kaist.ac.kr}}).
J. Park (Corresponding author) is with the School of Electrical and Electronic Engineering, Yonsei University, Seoul, South Korea (e-mail: {\texttt{jhpark@yonsei.ac.kr}}). 
N. Lee is with the School of Electrical Engineering, Korea University, Seoul, South Korea (e-mail: {\texttt{namyoon@korea.ac.kr}}).
A. Alkhateeb is with the School of Electrical, Computer and Energy Engineering, Arizona State University, Tempe, AZ, USA (e-mail: {\texttt{alkhateeb@asu.edu}}).

This research was supported in part by the MSIT(Ministry of Science and ICT), Korea, under the ITRC(Information Technology Research Center) support program(IITP-2023-RS-2023-00259991) supervised by the IITP(Institute for Information $\&$ Communications Technology Planning $\&$ Evaluation) and in part by by the National
Research Foundation of Korea (NRF) grant funded by the Korea Government (MSIT) under Grant 2022R1F1A1074391.
}
}

\maketitle \setcounter{page}{1}

\begin{abstract} 
Integrated sensing and communication (ISAC) is widely recognized as a fundamental enabler for future wireless communications. 
In this paper, we present a joint communication and radar beamforming framework for maximizing a sum spectral efficiency (SE) while guaranteeing desired radar  performance with imperfect channel state information (CSI) in multi-user and multi-target ISAC systems.
To this end, we adopt either a radar transmit beam mean square error (MSE) or receive signal-to-clutter-plus-noise ratio (SCNR) as a radar performance constraint of a sum SE maximization problem.
To resolve inherent challenges such as non-convexity and imperfect CSI, we  reformulate the problems and identify first-order optimality conditions for the joint radar and communication beamformer.
Turning the condition to a nonlinear eigenvalue problem with eigenvector dependency (NEPv), we develop an alternating method which finds the joint beamformer through power iteration and a Lagrangian multiplier through binary search.
The proposed framework encompasses both the radar metrics and is robust to channel estimation error with low complexity. 
Simulations validate the proposed methods.
In particular, we observe that the MSE and SCNR constraints exhibit  complementary performance depending on the operating environment, which manifests the importance of the proposed comprehensive and robust optimization framework.
\end{abstract}

\section{Introduction}
The upcoming era of wireless communication networks is poised to accommodate substantial data transmission speeds, top-notch wireless connectivity for a multitude of devices, and remarkably precise and resilient sensing capabilities~\cite{saad2019vision,liu2022integrated}.
Motivated by this, the concept of integrated sensing and communication (ISAC), which unifies signal processing techniques and hardware infrastructure between sensing and communication systems, is widely acknowledged as a pivotal technology for the realization of future wireless networks~\cite{rahman2019framework, liu2020joint, que2023joint}.
The integration of radar and communications on a unified platform holds the potential to reduce platform costs, facilitate spectrum sharing, and enhance performance through the synergy between radar and communications. 
Perceiving these promising advantages, we investigate a  multiple-input and multiple-output (MIMO) ISAC system and propose a novel MIMO-ISAC beamforming framework.


\subsection{Related Works}
In early studies, a system in which a separately located radar base station (BS) and cellular BS share spectrum with cooperation was mainly considered. 
In \cite{saru:jasc:12}, a simple opportunistic spectrum sharing method was proposed by exploiting radar rotation information, showing the potential of radar-communication coexistence.
Assuming the similar setup to \cite{saru:jasc:12}, in \cite{raymond:jsac:16}, a power control method was developed and it was revealed that the power control allows very short spatial separation between the radar and communication BSs. 
Considering a MIMO radar system, a null space projection approach was proposed in \cite{sodagari:gc:12} for minimizing a negative impact of a radar signal to communication systems. Incorporating the impact of the reflected radar signal from clutters, a beamforming optimization problem was tackled in \cite{qian:tsp:18}. In this work, the importance of considering the channel estimation error was demonstrated from simulations.
In \cite{liu:wcl:17}, a robust beamforming technique was developed to cope with the imperfect knowledge on channel state information at a transmitter (CSIT) for coexistence of a radar and a cellular BS.

Despite the benefits obtained from the coexistence of radar and communication, 
simply sharing the spectrum with separated locations has fundamental limitations. 
That is to say, a significant amount of overheads is caused in exchanging information between radar  and  cellular infrastructures for  cooperation.  
To overcome this, not only sharing the spectrum, but also unifying infrastructures of radar and communication has drawn significant attentions; leading to the emergence of an integrated sensing and communication (ISAC) system \cite{paul:access:17, zheng:spmag:19,kumari:tvt:18, dingyou:spmag:20,liu:jsac:22}. 
Nonetheless, several challenges newly have arisen for realizing the ISAC system, mainly caused by the difference between the primary goals of the two different applications~\cite{liu:jsac:22}.

To resolve the difficulties in realizing the unified system, extensive investigation has been performed for the ISAC system design. 
In \cite{sturm:proc:11}, it studied the sensing performance of various waveform including orthogonal frequency-division multiplexing (OFDM) and direct sequence spread spectrum and showed that OFDM  offers many advantages regarding the performance of the radar application. 
Information-theoretical approaches for analyzing and designing the unified system were also proposed by integrating not only the hardware infrastructure but also the performance metrics of the sensing and the communications  \cite{chiri:tsp:16, choi2023information}.
Using the unified information-theoretical radar and communication performance metrics, transmit waveform design was tackled in \cite{liu:commlett:17}.

The aforementioned ISAC studies \cite{sturm:proc:11,moghaddasi:access:16, chiri:tsp:16, liu:commlett:17, keskin:tsp:21} mainly assumed a single-antenna system.
An essential issue for achieving high efficiency in ISAC systems is forming multiple beams carefully by considering multiple communication users and radar targets; which propelled the use of a MIMO-ISAC system.
In \cite{hassanien:tsp:16}, an information embedding technique wherein communication signals are conveyed into MIMO radar beams by controlling their side-lobes was proposed assuming a line-of-sight channel. 
In addition, index-modulation type transmission schemes for delivering communication signals through MIMO radar waveform were proposed in \cite{ma:jstsp:21, huang:tsp:20, ma:tvt:21, baxter:tsp:22}.

For MIMO-ISAC systems, ISAC transmit beamforming optimization has been actively explored in the literature. 
In \cite{liu:twc:18, liu:tsp:18}, joint beamforming optimization methods were proposed. 
The key idea is employing the manifold optimization technique, by which the mean squared error (MSE) between the desired beampattern and the current beampattern is minimized while satisfying a power constraint. 
In \cite{liu:twc:18, liu:tsp:18}, communication symbols are used not only for downlink user service, but also for radar target sensing; yet this is sub-optimal compared to a case that uses designated radar sensing symbols \cite{hua2023optimal}. 
Addressing this limitation, in \cite{liu:tsp:20}, a joint beamforming optimization framework was proposed for a MIMO-ISAC system wherein radar sensing symbols are designated.
A similar problem was also considered in \cite{hua2023optimal}, assuming that  each communication user is cable of eliminating radar sensing signals.
In \cite{chen:tsp:21}, a fairness-profile optimization problem was addressed for joint precoding and radar signaling, in which a difference between peak and side lobe of radar beams is controlled with communication signal-to-interference-plus-noise ratio (SINR) constraints. 
In \cite{xu:access:20}, a sum rate maximization problem was formulated with radar transmit beam at target direction as a penalty term and solved by alternating methods. 
Considering a dirty paper coding scheme, joint ISAC beamforming problems were also tackled in \cite{liuX:jsac:22} by leveraging the uplink-downlink duality.

Beyond conventional downlink setups, various scenarios of ISAC systems has also been explored recently. 
Incorporating a concept of physical layer security into a MIMO-ISAC system, secure transmission methods were proposed in \cite{su:twc:21, ren2023robust}. 
Employing advanced multiple access technique such as rate-splitting multiple access (RSMA)  or non-orthognal multiple access (NOMA) into a MIMO-ISAC system, RSMA-based MIMO-ISAC system \cite{xu:jstsp:21} and  NOMA-based MIMO-ISAC system \cite{Mu:jsac:22} were investigated.

Previous research endeavors on ISAC systems have adeptly elucidated the merits of integrating these systems while furnishing cutting-edge optimization techniques aimed at enhancing both radar and communication performance~\cite{liu:tsp:20,xu:access:20}. 
These investigations have identified that a shared-antenna ISAC system surpasses its separate-antenna counterpart, and underscored the desirability of leveraging dedicated radar signals to fully exploit the available spatial degrees of freedom (DoF). 
Consequently, the development of ISAC beamforming strategies tailored to shared-antenna systems with designated radar signals, especially under realistic scenarios such as imperfect CSIT, multiple users, and multiple targets, is regarded as a pertinent research imperative; yet it has not been completely addressed.
For instance, in~\cite{su:twc:21, ren2023robust}, robust transmission strategies using imperfect CSIT were proposed, primarily targeting scenarios involving a single target or  a single user, respectively. 
There exist some prior works that considered multi-user scenarios~\cite{xu:access:20,liu:tsp:20} and multi-target settings~\cite{liu:tsp:20}; yet they predominantly assumed perfect CSIT. 
In response to this research gap, we propose a comprehensive framework for joint and robust beamforming optimization, harnessing dedicated radar signals in the shared-antenna ISAC systems that support multiple users and cater to multiple targets under the assumption of the imperfect CSIT.

\subsection{Contributions}


In this paper, we propose a robust and joint beamforming optimization framework for MIMO-ISAC systems to maximize the sum spectral efficiency (SE) while achieving the desired radar performances. 
The proposed framework is designed for multi-user and multi-target scenarios, and comprehensive radar metrics with designated radar signals under the imperfect CSIT assumption.
Our main contributions are summarized as: 
\begin{itemize}
    \item 
    We formulate two communication-oriented ISAC beamforming problems, each of which considers distinct radar  metrics: radar transmit beam MSE and radar receive signal-to-clutter-plus-noise ratio (SCNR). 
    These problems involve maximizing the sum SE subject to either the MSE or the SCNR constraint, by which our problem incorporates the transmit and receive perspectives of radar probing signals, respectively. 
    However, due to the non-convex nature of these optimization problems and the presence of the imperfect CSIT, addressing these problems are challenging. 
    To resolve the challenge, we propose a joint beamforming optimization framework. Our method is featured by 
    1) capability of accommodating both radar metrics (the MSE and the SCNR), 2) robustness against the imperfect CSIT, and 3) efficiency in achieving strong communication and sensing performances. 
    \item To this end, we first reformulate the problems by deriving a lower bound on the average SE in terms of the channel estimation error. 
    This reformulation allows us to leverage the channel estimation error covariance in optimization. 
    Then we obtain a tractable form of the problems by representing the objective function and constraints as a function of a vectorized joint communication and radar beamformer.
    Subsequently, we compute the Lagrangian function of each problem and derive a stationarity condition of the  Karush–Kuhn–Tucker (KKT) condition with respect to the vectorized beamformer to identify local optimal points for a given Lagrangian multiplier. 
    \item In pursuit of finding the best stationary point that maximizes the Lagrangian function, we interpret the obtained stationarity condition as a nonlinear eigenvalue problem with eigenvector dependency (NEPv). 
    In the NEPv, the eigenvector and eigenvalue correspond to the vectorized beamformer and Lagrangian, respectively. 
    This observation indicates that the principal eigenvector of the NEPv is equivalent to the best stationary point. 
    Accordingly, for the given Lagrangian multiplier, we propose a generalized power iteration (GPI)-based algorithm to find the principal eigenvector. Given the beamforming vector, a binary search-based algorithm is used to update a proper Lagrangian multiplier that meets the radar constraint. 
    The beamforming vector and the Lagrangian multiplier are updated in an alternating manner until convergence.
    As a special case, we also put forth radar-only beam design algorithms by using our proposed framework.
    \item Through simulations, we demonstrate that the proposed algorithms outperform other benchmark algorithms by offering a better trade-off between radar and communication performance and more robust performance under the imperfect CSIT. 
    In addition to the performance gains, our method also sheds lights on crucial design guidelines for MIMO-ISAC, that could not be extracted through existing methods. 
    For example, it becomes evident that careful selection of the radar metrics is desirable: the considered radar metrics exhibit a complementary relationship to each other depending on the operating environment. 
    This observation underpins the importance of the proposed comprehensive  optimization framework.
\end{itemize}

\textit{Notation}:
$\text{a}$ is a scalar, ${\bf{a}}$ is a vector and ${\bf{A}}$ is a matrix.
Superscripts $(\cdot)^{\sf T}$, $(\cdot)^{\sf H}$, and $(\cdot)^{-1}$ are matrix transpose, Hermitian, and inversion, respectively.
$\mathbb{E}[\cdot]$ and $\text{tr}(\cdot)$ represent  expectation operation and trace of a matrix, respectively.
${\bf{I}}_N$ is an $N \times N$ identity matrix.
${\bf 0}_{N\times M}$ is the  $N \times M$ zero matrix.
We denote $\otimes$ as kronecker product, and ${\bf e}^{N}_n$ as an $N\times 1$ standard basis vector whose $n$th element is one. 
${\rm diag}({\bf a})$ is a diagonal matrix with ${\bf a}$ on its diagonal elements.
$\mathcal{CN}(m,\sigma^2)$ is a complex Gaussian distribution with mean $m$ and variance $\sigma^2$.
\section{System Model}


\subsection{Signal Model}
We consider a MIMO-ISAC system, where the BS is equipped with a ISAC transmitter and a radar receiver operating on a single-hardware platform with $N$ transmit and receive antennas.
The BS performs two disparate functions: 1) serving $K$ single antenna communication users and 2) forming radar beams for sensing potential targets. 
We refer to this BS as ISAC-BS. 
These two functionalities simultaneously share time-frequency resources and transmit antenna arrays. 
As assumed in~\cite{keskin:tsp:21}, we consider complete decoupling between transmit and receive antennas, ensuring that the radar receiver remains free from self-interference.
To maximally exploit the spatial DoF provided by the MIMO system, the BS uses dedicated radar signals by superimposing them with communication symbols.
Accordingly, the BS transmits the signal ${\bf{x}} \in \mathbb{C}^{N }$ defined as 
\begin{align}
    {\bf{x}} = {\bf{F}}_{\text C} {\bf{s}}_{\text C} + {\bf{F}}_{\text R} {\bf{s}}_{\text R} = [{\bf{F}}_{\text C}, {\bf{F}}_{\text R}] \left[{\bf{s}}_{\text C}^{\sf T}, {\bf{s}}_{\text R}^{\sf T} \right]^{\sf T} = {\bf{F}} {\bf{s}}, 
\end{align}
where ${\bf{s}}_{\text R} = [s_{{\rm R},1},\dots,s_{{\rm R},M}]^{\sf T} \in \mathbb{C}^M$ is a radar waveform signal vector, ${\bf{F}}_{\text R} = \left[{\bf{f}}_{{\text R},1}, \cdots, {\bf{f}}_{{\text R},M} \right] \in \mathbb{C}^{N \times M}$ is a radar beamforming matrix, ${\bf{s}}_{\text C} = [s_{{\rm C},1},\dots,s_{{\rm C},K}]^{\sf T} \in \mathbb{C}^K$ is a communication message symbol vector, ${\bf{F}}_{\text C} = \left[{\bf{f}}_{{\text C},1}, \cdots, {\bf{f}}_{{\text C},K}\right] \in \mathbb{C}^{N \times K}$ is a communication beamforming matrix,
${\bf F} = [{\bf F}_{\rm C}, {\bf F}_{\rm R}] \in \mathbb{C}^{N\times (K+M)}$ is a concatenated matrix composed of ${\bf F}_{\rm C}$ and ${\bf F}_{\rm R}$, and ${\bf s} = [{\bf{s}}_{\rm C}^{\sf T}, {\bf{s}}_{\rm R}^{\sf T} ]^{\sf T} \in \mathbb{C}^{(K+M)\times 1}$ is a concatenated vector composed of ${\bf{s}}_{\rm C}$ and  ${\bf{s}}_{\rm R}$.
Here, $s_{{\rm R},m}$ is a radar-dedicated symbol introduced for a sensing beam design purpose only,  ${\bf f}_{{\rm R},m}$ is the corresponding sensing beamforming vector, $s_{{\rm C},k}$ is the communication symbol for user $k$, and ${\bf f}_{{\rm C}, k}$ is its corresponding beamforming vector. 

We further assume that the radar waveform signal ${\bf{s}}_{\text R}$ and the communication symbols ${\bf{s}}_{\text C}$ are zero-mean and statistically independent to each other, i.e., $\mathbb{E}\left[{\bf{s}}_{\text R} {\bf{s}}_{\text C}^{\sf H}\right] = {\bf{0}}_{M\times K}$, with $\mathbb{E}\left[{\bf{s}}_{\text R} {\bf{s}}_{\text R}^{\sf H}\right] = P\cdot{\bf{I}}_M$, and $\mathbb{E}\left[{\bf{s}}_{\text C} {\bf{s}}_{\text C}^{\sf H}\right] = P\cdot {\bf{I}}_K$, where $P$ is the transmit power. 
Fulfilling a transmit power constraint, we assume $\|{\bf{F}}_{\text R}\|_{\sf F} + \|{\bf{F}}_{\text C} \|_{\sf F} \le 1$ without loss of generality. 




\subsection{Channel Model}

The received signal $y_k$  at user $k$  is given by 
\begin{align} 
    \label{eq:y_k}
    y_k 
    &= {\bf{h}}_k^{\sf H} {\bf{f}}_{{\text C},k} {{s}}_{{\text C},k} \!+\! \sum_{i \neq k, i = 1}^{K} {\bf{h}}_k^{\sf H} {\bf{f}}_{{\text C},i} {{s}}_{{\text C},i}\! +\! \sum_{j = 1}^{M} {\bf{h}}_k^{\sf H} {\bf{f}}_{{\text R},j} {{s}}_{{\text R},j} \!+\! n_k,
\end{align}
\!\!where ${\bf{h}}_k \in \mathbb{C}^{N}$ is the channel vector from the BS and user $k$, and $n_k \sim \mathcal{CN}(0, \sigma^2)$ is the additive white Gaussian noise (AWGN). 
In \eqref{eq:y_k}, the second term on the right side indicates the aggregated interference from other communication user signals, and third term is the aggregated interference from the radar-sensing waveform signals. 

For each channel vector, we assume ${\bf{h}}_k \sim \mathcal{CN}({\bf 0}_{K\times 1}, {\bf{K}}_{k})$, i.e., it is distributed as Gaussian with a specific spatial covariance matrix ${\bf{K}}_k$.
Upon this channel model, ${\bf{h}}_k$ is given as
 \begin{align} \label{eq:channel_model}
 	{\bf{h}}_k = {\bf{U}}_k {\bf \Lambda}_k^{\frac{1}{2}} {\tilde{\bf{h}}}_k,
 \end{align}
by following the Karhunen–Loeve transformation \cite{adhi:tit:13, jiang:twc:15}.
In \eqref{eq:channel_model}, ${\bf \Lambda}_k \in \mathbb{C}^{r_k \times r_k}$ is a diagonal matrix that contains the non-zero eigenvalues of ${\bf{K}}_k$, ${\bf{U}}_k \in \mathbb{C}^{N \times r_k}$ is a collection of the eigenvectors of ${\bf{K}}_k$ corresponding to ${\bf \Lambda}_k$, and $\tilde{\bf{h}}_k \in \mathbb{C}^{r_k}$ is a independent and identically distributed (IID) channel vector whose element is drawn from $\mathcal{CN}(0,1)$. 
We note that our channel model encompasses most of the widely-used spatially correlated MIMO channel models as explained in the following remark:
\begin{remark}
    \normalfont (Embracing Other Channel Models) For instance,  considering a well-known multi-path channel model \cite{han:twc:19, xie:twc:18}, we represent 
    \begin{align} \label{eq:multipath}
        {\bf{h}}_k =  \sum_{\ell = 1}^{L_k} \alpha_{\ell, k} {\bf{a}}(\theta_{\ell,k}), 
    \end{align}
    where ${\bf{a}}(\theta_{\ell, k})$ is an array steering vector with angle of departure (AoD) $\theta_{\ell, k}$ and $\alpha_{\ell, k}$ is a complex path gain drawn from Gaussian $\mathcal{CN}(0, \sigma^2_{\ell, k})$. 
    Because AoDs are determined by certain propagation geometries, which are typically slowly varying \cite{adhi:tit:13}, the covariance matrix is computed as 
    \begin{align} \label{eq:mp_cov}
    \mathbb{E} \left[{\bf{h}}_k {\bf{h}}_k^{\sf H} \right] = \sum_{\ell = 1}^{L_k} \sigma_{\ell, k}^{2} {\bf{a}}(\theta_{\ell, k}) {\bf{a}}(\theta_{\ell, k})^{\sf H} = {\bf{K}}_k = {\bf{U}}_k {\bf{\Lambda}}_k {\bf{U}}_k^{\sf H}.
    \end{align}
    Since \eqref{eq:multipath} is interpreted as a linear sum of IID Gaussian random variables, this can be easily expressed by \eqref{eq:channel_model} with \eqref{eq:mp_cov}. 
    Note that this connection was also well presented in \cite{yin:jsac:13}. 
\end{remark}


Now we describe the CSIT error model. 
We first clarify that our framework is applicable in any type of CSIT imperfection. 
However, we specifically focus on a CSIT imperfection model relevant to frequency division duplex (FDD) MIMO systems. 
This is because a FDD system is more favorable for the considered ISAC scenario compard to a time division duplex (TDD) system due to its orthogonal nature of the echo signal to the uplink user signal.
In FDD systems, the obtained CSIT in the BS is modeled as
    \begin{align} \label{eq:csit error}
        \hat {\bf{h}}_k = {\bf{h}}_k -  {\pmb{\phi}}_k,
    \end{align}
where $ {\pmb{\phi}}_k$ is the CSIT error vector.  
Since the CSIT error is mostly caused by quantization for uplink feedback \cite{wagner:tit:12}, the statistics of $ {\pmb{\phi}}_k$ is mainly determined by the used quantization method. 
In this paper, we consider entropy-coded scalar quantization (ECSQ), which quantizes each component of the channel vector by applying the Karhunen–Loeve transformation. 
As in \cite{jiang:twc:15} the covariance of $ {\pmb{\phi}}_k$ under ECSQ is 
\begin{align} \label{eq:phi_k}
    {\bf{\Phi}}_k = \mathbb{E}[{\pmb{\phi}}_k{\pmb{\phi}}^{\sf{H}}_k] = {\bf{U}}_k \kappa^2 {\bf{\Lambda}}_k {\bf{U}}_k^{\sf H}
\end{align}
while $\hat {\bf{h}}_k$ and ${\pmb{\phi}}_k$ are uncorrelated, and $\mathbb{E}[{\pmb{\phi}}_k] = 0$ since $\mathbb{E}[{\bf{h}}_k] = 0$ and $\mathbb{E}[\hat {\bf{h}}_k]=0$ \cite{gray:tit:98}. 
In \eqref{eq:phi_k}, $\kappa \in [0,1]$ indicates the accuracy of quantization, wherein smaller $\kappa$ means accurate quantization. 
Correspondingly, $\kappa = 0$ means that enough quantization bits are used, making the CSIT error negligible. 


Now we present our CSIT error treatment technique. 
In general, the accurate distribution on ${\pmb{\phi}}_k$ is not tractable.
Hence it is not feasible to obtain the exact mutual information expression regarding ${\pmb{\phi}}_k$. 
To resolve this challenge, we treat ${\pmb{\phi}}_k$ as Gaussian distribution, which leads to a lower bound on the achievable spectral efficiency when it is uncorrelated to the desired signal \cite{hassibi2003much}.
This is more clearly explained in Section \ref{sec:reformulation}. 
We also note that many prior works employed this approach to handle the CSIT error in a tractable manner \cite{jiang:twc:15, wagner:tit:12}. 
Additionally, in a high resolution quantization case, approximating quantization error as Gaussian is known to be convenient for parameter estimation or quantization performance characterization as studied in \cite{marco:tit:05, kipnis:tit:21}.

\begin{remark} \normalfont (CSIT Error in VQ)
    Vector quantization (VQ) is another popular quantization method \cite{Park:twc, jindal:tit:06}. 
    In VQ, we often model the CSIT error as a bounded error. Specifically,  we assume $ \|{\pmb{\phi}}_k \|^2 \le \delta_k$ in \eqref{eq:csit error}, where $\delta_k$ is determined as a function of the used feedback bits $B$, e.g., $ \delta_k\sim 2^{-\frac{B}{N-1}}$. 
    In this case, one can employ a conservative approach to handle the CSIT error; denoting the uncertainty region of the true channel as $\mathbb{H}_k = \{{\bf{h}}_k | {\bf{h}}_k = \hat {\bf{h}}_k + {\pmb{\phi}}_k, \|{\pmb{\phi}}_k\| \le \delta_k \}$, 
    the conservative approach first obtains ${\bf{h}}_k$ for $k = 1, \cdots, K$ that leads to the worst case of the considered problem.
    Then it finds a solution which maximizes the objective function for the worst case scenario. 
    By doing this, we can achieve a robust solution. 
    In a sense that the conservative approach pursues to optimize under the worst case scenario, our Gaussian treatment and the conservative approach share a common thread.
\end{remark}

\subsection{Radar Model}

In many instances, successful sensing necessitates a direct line-of-sight (LoS) between the sensor and the target object. 
In conventional radar applications, signals bouncing off non-target objects are known as clutter.
Unlike the communication systems which take advantage of the multiple paths of channels, the clutter is considered detrimental for sensing accuracy and requires mitigation.
Involving the clutter, the received radar echo signal is modeled as \cite{liu:jsac:22}
\begin{align}
    \nonumber
    {\bf y}_{\rm R} &= \sum_{i=1}^S\beta_i^{\rm tar} {\bf a}_{\rm r}(\theta^{\rm tar}_i){\bf a}_{\rm t}^{\sf H}(\theta^{\rm tar}_i){\bf x} + \sum_{j=1}^C\beta_j^{\rm cl} {\bf a}_{\rm r}(\theta^{\rm cl}_j){\bf a}_{\rm t}^{\sf H}(\theta^{\rm cl}_j){\bf x} + {\bf n}_{\rm R}
    \\\label{eq:y_R}
    &= {\bf G}_{\rm tar}{\bf x} + {\bf G}_{\rm cl}{\bf x} + {\bf n}_{\rm R},
\end{align}
where $S$ is the number of targets, $C$ is the number of clutters, ${\bf a}_{\rm t}(\theta)$ is the transmitter array response vector to the direction $\theta$, ${\bf a}_{\rm r}(\theta)$ is the receiver array response vector to the direction $\theta$, $\theta^{\rm tar}_i$ is the  direction of the $i$th target,  $\theta^{\rm cl}_j$ is the direction of the $j$th clutter, $\beta_i^{\rm tar}$ is the reflection coefficient of the $i$th target, $\beta_j^{\rm cl}$ is the reflection coefficient of the $j$th  clutter, and ${\bf n}_{\rm R} \sim \mathcal{CN}({\bf  0}_{N \times 1}, \sigma^2_{\rm R}{\bf I}_N)$ is the AWGN vector.
Here ${\bf G}_{\rm tar} = \sum_{i=1}^S\beta_i^{\rm tar} {\bf a}_{\rm r}(\theta^{\rm tar}_i){\bf a}_{\rm t}^{\sf H}(\theta^{\rm tar}_i)$ and ${\bf G}_{\rm cl} = \sum_{j=1}^C\beta_j^{\rm cl} {\bf a}_{\rm r}(\theta^{\rm cl}_j){\bf a}_{\rm t}^{\sf H}(\theta^{\rm cl}_j) $.
The reflection coefficient is a function of radar cross section, target distance, and carrier frequency \cite{braun2014ofdm}. It is noted that the radar echo channel model for each target has a single propagation path while the communication channel model considers multiple propagation paths that can include propagation paths due to clutter reflection.

We consider that the reflection coefficients of the targets $\beta_i^{\rm tar}$ and clutters $\beta_j^{\rm cl}$  and  the directions of targets $\theta_i^{\rm tar}$ and clutters $\theta_j^{\rm cl}$ are known at the BS.
In other words, there can be an initial beam sweeping phase prior to the data transmission and radar sensing phase considered in this paper. 
In this initial phase, basic target information such as distance, angle, and radar cross section can be roughly estimated to use for forming a radar tracking beam in the next phase.
Accordingly, one of the possible application scenarios of our proposed algorithms is such a data transmission and tracking phase, but not limited to this scenario.
In addition, when the clutters are fixed, e.g. buildings, light poles, trees, etc., $\theta_j^{\rm cl}$ and $\beta_j^{\rm cl}$ are almost time-invariant, which further allows us to use the information.

\subsection{Performance Metrics}

\subsubsection{Communication Performance Metric}
We use a sum SE as a performance  metric. 
The SINR at user $k$ is given as
\begin{align} \label{eq:sinr}
    \gamma_k = \frac{\left|{\bf{h}}_k^{\sf H}{\bf{f}}_{{\text C},k}\right|^2}{\sum_{i \neq k, i = 1}^{K} \left|{\bf{h}}_k^{\sf H}{\bf{f}}_{{\text C},i}\right|^2 + \sum_{j = 1}^{M}\left|{\bf{h}}_k^{\sf H}{\bf{f}}_{{\text R},j}\right|^2 + \frac{\sigma^2}{P}}.
\end{align}
In \eqref{eq:sinr}, we observe that the interference includes both {\it{i}}) the inter-user interference and {\it{ii}}) the radar waveform interference.
The SE achieved at user $k$ is presented as
\begin{align}
    \label{eq:rate}
    R_k =  \log_2 \left(1 + \gamma_k \right).
\end{align}


\subsubsection{Radar Performance Metric}

We consider two  radar performance metrics: radar transmit beam pattern MSE and radar receive SCNR, and formulate problems with each metric. 

{\bf Mean Squared Error (MSE):} As in the prior works \cite{hua2023optimal, liu:twc:18, liu:tsp:20}, the MSE between the transmit beam pattern and the desired beam pattern is often considered as a key radar performance metric. 
Specifically, assuming that the ISAC-BS operates in a narrowband and the propagation paths to  radar sensing targets are LoS, the baseband signal at $\theta$ is 
\begin{align} \label{eq:radar_signal}
    z(\theta) = {\bf{a}}_{\rm t}^{\sf H}(\theta) {\bf{x}}.
\end{align}
We note that the ISAC-BS is able to exploit not only the radar waveform signals ${\bf{s}}_{\rm R}$, but also the communication signals ${\bf{s}}_{\rm C}$ in forming radar sensing beam patterns.
With \eqref{eq:radar_signal}, the radar beam power for an angular direction $\theta$ is characterized as
\begin{align} \label{eq:radar_beam}
    B({\bf F};\theta) &= \mathbb{E}\left[|z(\theta)|^2\right] = P \cdot {\bf{a}}_{\rm t}^{\sf H}(\theta) {\bf{F}} {\bf{F}}^{\sf H} {\bf{a}}_{\rm t}(\theta). 
\end{align}
Now we define the MSE between the formed beam patterns and the given desired beam patterns as follows:  
\begin{align}
    \label{eq:MSE}
    {\text{MSE}}_{\text R}({\bf F}) = \frac{1}{L} \sum_{\ell = 1}^{L} |P \cdot d(\theta_\ell) - B({\bf F};\theta_{\ell}) |^2,
\end{align}
where $d(\theta)$ is the given normalized desired transmit beam mask for the direction $\theta$, and $L$ is the number of sampled angular grids. 
As ${\text{MSE}}_{\text R}$ decreases, the actual radar beam patterns become similar to the desired beam patterns. 
Accordingly, the advantage of using  ${\text{MSE}}_{\text R}$ as a radar performance metric is that the transmit beam pattern can be directly designed to match the desired beam pattern. 
In this regard, it is particularly useful to design beams with  multiple target angles.

{\bf Signal-to-Clutter-plus-Noise Ratio (SCNR):} The limitation of using ${\text{MSE}}_{\text R}$ \eqref{eq:MSE} is that it is hard to incorporate the effect of the undesirable echo signals, i.e., clutters. 
To resolve this problem, the SCNR can be used as a radar performance metric when clutters are non-negligible.
For instance, the SCNR metric can be effective for a radar tracking phase in which the initial target positions and clutter positions are available.
From \eqref{eq:y_R}, the SCNR is computed as \cite{cui2013mimo, liu:jsac:22}
\begin{align}
    \label{eq:scnr}
    \gamma_{\rm R}({\bf F}) \!=\! \frac{\mathbb{E}[\|{\bf G}_{\rm tar}{\bf x}\|_{\sf F}]}{\mathbb{E}[\|{\bf G}_{\rm cl}{\bf x}\|_{\sf F}  \!+\! \|{\bf n}_{\rm R}\|_{\sf F}]} 
    \!=\! \frac{{\rm Tr}\left({\bf G}_{\rm tar}{\bf F}{\bf F}^{\sf H}{\bf G}_{\rm tar}^{\sf H}\right)}{{\rm Tr}\left({\bf G}_{\rm cl}{\bf F}{\bf F}^{\sf H}{\bf G}_{\rm cl}^{\sf H}\right)\! +\! \frac{N}{P}\sigma^2_{\rm R}}.
\end{align}
Using the SCNR in \eqref{eq:scnr} allows to take into account the effect of the clutters, 
whereas direct beam pattern design cannot be performed, which is different from using the MSE.
Consequently, beam patterns for multiple targets may not be suitable with the SCNR as it considers the sum power from all targets, not individual power. 
Therefore, the MSE and SCNR can be viewed as complementary to each other.
This conjecture will be confirmed in Section~\ref{sec:sim}.

\begin{remark} \normalfont (Discussion on Sensing Metrics)
In some prior works, the Cramer-Rao bound (CRB) was adopted as a sensing metric  \cite{liu:tsp:22}.
The CRB and the MSE (or the SCNR) serve different purposes in sensing system design and performance evaluation. 
The CRB is useful in providing a theoretical guideline to fundamental estimation performances, given the exact point target angle. 
However, it cannot incorporate various ISAC system design aspects beyond estimation performance: for example, mitigating the probing signal power to certain directions for interference mitigation or avoiding clutters. 
Furthermore, the exact knowledge on the point target angle (or the corresponding statistics) is needed to be given to compute the CRB \cite{liu:tsp:22}, which is often impractical. 
For this reason, the CRB is more suited for static target detection or developing and evaluating  detection algorithms with given sensing beam.

On the contrary to that, adopting the MSE, it is possible to reduce the probing signal power in specific directions by adjusting the desired  beam mask. 
In addition, using the MSE can adapt a case in which the BS does not have the exact knowledge on the point target angle; 
if there are some uncertainties on the point target angle, we can widen the radar beam width, so as to collect the reflected signal over a broader range. 
Moreover, using the MSE, the target estimation performance can be guaranteed implicitly to approach well-designed radar beam pattern \cite{yin:commlett:22}. 
Using the SCNR, it is feasible to achieve high sensing performance by avoiding clutters. 
Due to these advantages, many prior ISAC studies considered the radar beam pattern MSE or the SCNR as a main sensing metric  \cite{liu:twc:18, liu:tsp:20, liu:jsac:22}. 
Correspondingly, the MSE and the SCNR are suited for initial sensing beam sweeping as well as target tracking beam generation. 
Extending our framework by incorporating the CRB still is interesting future work. 
\end{remark}


\section{Proposed Optimization Framework}

In this section, we first formulate the ISAC beamforming problem with the considered metrics and then put forth our unified framework to solve.
To this end, we  reformulate the objective function and constraints into tractable forms. 
Leveraging the reformulation, we study the first-order optimality condition, and then develop an efficient algorithm to find a superior local optimal point. 

\subsection{Problem Formulation}

Considering communication-oriented ISAC systems, our main design goal is to maximize the sum SE under a constraint that the radar beam pattern MSE is smaller than the predefined threshold $T_{\rm mse}$:
\begin{align} \label{eq:prob_formulation}
    \mathop{\text {maximize}}_{{\bf{F}}}\;\; &  R_{\Sigma}({\bf F}) = \sum_{k = 1}^{K} R_k({\bf F})
    \\
    {\text {subject}\;\text {to}} \;\; & {\rm MSE}_{\rm R}({\bf F}) \le T_{\rm mse}, 
    \\\label{eq:prob_PowerConst}
    & {\rm Tr}({\bf F}{\bf F}^{\sf H}) \leq  1,
\end{align}
where \eqref{eq:prob_PowerConst} indicates the transmit power constraint.
Similarly, with the SCNR performance metric, we formulate the problem:
\begin{align} \label{eq:prob_formulation_scnr}
    \mathop{\text {maximize}}_{{\bf{F}}}\;\; & R_{\Sigma}({\bf F}) = \sum_{k = 1}^{K} R_k({\bf F}) 
    \\
    {\text {subject}\;\text {to}} \;\; & \gamma_{\rm R}({\bf F}) \ge T_{\rm scnr}, 
    \\
     &{\rm Tr}({\bf F}{\bf F}^{\sf H}) \leq  1.
\end{align}
The problems are non-convex, and due to the assumption of the imperfect CSIT, the sum SE $R_{\Sigma}$ needs to be handled to fully utilize the estimated channel and error covariance matrix.

\subsection{Problem Reformulation with MSE Constraint} \label{sec:reformulation}

We recall that the BS is only able to use the estimated CSIT $\hat {\bf{h}}_k$ and its error covariance matrix ${\bf \Phi}_k$. 
Accordingly,
it is necessary to obtain a proper SE expression instead of $R_{\Sigma}({\bf F})$.
To this end, we consider the average SE in terms of the CSIT estimation error \cite{joudeh:16:tcom} which is computed as  
\begin{align}
    \label{eq:avg_se}
    &\bar R_k({\bf F})  = \mathbb{E}_{\pmb{\phi}_k}[R_k({\bf F}) ]
    \\\nonumber
    &\!\!\!=\! \mathbb{E}_{{\pmb{\phi}}_k} \!\!\left[\log_2\left(\!1 \! +\! \frac{\left|\left( \hat {\bf{h}}_k \!+\! {\pmb{\phi}}_k \right)^{\sf H}{\bf{f}}_{{\text C},k}\right|^2}{\sum_{i \neq k}^{K}\! \left|\left( \hat {\bf{h}}_k \!+\! {\pmb{\phi}}_k \right)^{\sf H}\!\!\!{\bf{f}}_{{\text C},i}\right|^2 \!\!\!+\! \sum_{j = 1}^{M}\!\left|\left( \hat {\bf{h}}_k \!+\! {\pmb{\phi}}_k \right)^{\sf H}\!\!\!{\bf{f}}_{{\text C},j}\right|^2 \!\!\!+\! \frac{\sigma^2}{P}} \!\right) \right].
\end{align}
To convert \eqref{eq:avg_se} into a tractable closed form, we first rewrite the signal model \eqref{eq:y_k}  with the CSIT error term as
\begin{align}
    \nonumber
    y_k 
    &= \hat {\bf{h}}_k^{\sf H} {\bf{f}}_{{\text C},k} {{s}}_{{\text C},k} + \sum_{i \neq k, i = 1}^{K} \hat {\bf{h}}_k^{\sf H} {\bf{f}}_{{\text C},i} {{s}}_{{\text C},i} + \sum_{j = 1}^{M} \hat{\bf{h}}_k^{\sf H} {\bf{f}}_{{\text R},j} {{s}}_{{\text R},j} 
    \\
    &\quad +  \underbrace{\sum_{ i' = 1}^{K} {\pmb{\phi}}_k^{\sf H} {\bf{f}}_{{\text C},i'} {{s}}_{{\text C},i'}+ \sum_{j' = 1}^{M} {\pmb{\phi}}_k^{\sf H} {\bf{f}}_{{\text R},j'}{{s}}_{{\text R},j'}}_{(a)} + n_k.
\end{align}

Since the CSIT error ${\pmb{\phi}}_k$ is uncorrelated with the desired signal, 
treating the CSIT error-related term in $(a)$ as Gaussian noise results in a lower bound on the mutual information expression. 
This leads to a lower bound of the SE $R_k$. 
Note that this is also aligned with the concept of generalized mutual information \cite{yoo2006capacity,medard2000effect,lapidoth2002fading,ding2010maximum}.
Accordingly, we have the lower bound of \eqref{eq:avg_se} as shown in \eqref{eq:avg_se_lb2} at the top of the next page,
\begin{figure*}
\begin{align} 
    \label{eq:avg_se_lb} 
    \bar R_k({\bf F})  &\ge  \mathbb{E}_{{\pmb{\phi}}_k} \left[\log_2\left(1 + \frac{\left|\hat {\bf{h}}_k^{\sf H}{\bf{f}}_{{\text C},k}\right|^2}{\sum_{i \neq k, i = 1}^{K} \left|\hat {\bf{h}}_k^{\sf H}{\bf{f}}_{{\text C},i}\right|^2 + \sum_{j = 1}^{M}\left|\hat {\bf{h}}_k^{\sf H}{\bf{f}}_{{\text R},j}\right|^2 + \sum_{i' = 1}^{K} \left|{\pmb{\phi}}_k^{\sf H} {\bf{f}}_{{\text C},i'} \right|^2 + \sum_{j' = 1}^{M} \left| {\pmb{\phi}}_k^{\sf H} {\bf{f}}_{{\text R},j'} \right|^2 + \frac{\sigma^2}{P}} \right) \right] 
    \\
    & \mathop{\ge}^{(b) } \log_2\left(1 + \frac{\left|\hat {\bf{h}}_k^{\sf H}{\bf{f}}_{{\text C},k}\right|^2}{\sum_{i \neq k, i = 1}^{K} \left|\hat {\bf{h}}_k^{\sf H}{\bf{f}}_{{\text C},i}\right|^2 + \sum_{j = 1}^{M}\left|\hat {\bf{h}}_k^{\sf H}{\bf{f}}_{{\text R},j}\right|^2 + \sum_{i' = 1}^{K}  {\bf{f}}_{{\text C},i'}^{\sf H} {\bf{\Phi}}_k {\bf{f}}_{{\text C},i'}  + \sum_{j' = 1}^{M}   {\bf{f}}_{{\text R},j'}^{\sf H} {\bf{\Phi}}_k {\bf{f}}_{{\text R},j'}+ \frac{\sigma^2}{P}} \right) 
    \\ \label{eq:avg_se_lb2}
    & = \bar R_k^{\sf lb}({\bf F}).
\end{align}
    \vspace{-1 em}
    \rule{\textwidth}{0.8pt}
\end{figure*}
where $(b)$ follows from Jensen's inequality and $\mathbb{E}[{\pmb{\phi}}_k {\pmb{\phi}}_k^{\sf H}] = {\bf{\Phi}}_k$. Since $\bar R_k^{{\sf{lb}}}$ is derived as a closed form, we use this as a new objective function in our optimization problem. 
$\bar R_k^{\sf lb}$ is a function of  $\hat {\bf{h}}_k$ and  ${\bf{\Phi}}_k$  that are available at the BS; thereby $\bar R_k^{\sf lb}$ can be fully used to design the beamforming matrix at the BS side. 

Let us denote $\bar {\bf{f}}= {\text{vec}}({\bf{F}}) \in \mathbb{C}^{N(K+M)}$, i.e., the vectorized beamforming variable, and assume $\|\bar {\bf{f}}\|^2= 1$ which implies signaling with maximum transmit power $P$.
Since the sum SE increases with the transmit power, this assumption aligns with the problem in maximizing the sum SE.
Then, we further reformulate $\bar R_k^{\sf lb}$ for more tractability as 
\begin{align}
    \label{eq:R_lb_rayleigh}
    \bar R_k^{\sf lb}(\bar {\bf f}) =  \log_2 \left( \frac{\bar{\bf{f}}^{\sf H}{\bf{B}}_k \bar {\bf{f}}}{\bar{\bf{f}}^{\sf H} 
    {\bf{C}}_k\bar{\bf{f}}} \right), 
\end{align} 
where ${\bf{B}}_k = {\bf I}_{(K+M)}\otimes \big(\hat{\bf{h}}_k \hat{\bf{h}}_k^{\sf H} \!+\! {\bf{\Phi}}_k\big) \!+\! \frac{ \sigma^2}{P} {\bf{I}}_{N(K+M)}$ and ${\bf{C}}_{k} = {\bf{B}}_{k} - {\rm diag}\left({\bf e}^{(K+M)}_k\right)\otimes \hat{ \bf{h}}_k \hat{\bf{h}}_k^{\sf H}.$

The next step is to reformulate the MSE constraint as a function of ${\bar {\bf f}}$.
The radar beam power in \eqref{eq:radar_beam} is rewritten as
\begin{align}
    B({\bf F};\theta) &= P \cdot {\bf{a}}_{\rm t}^{\sf H}(\theta) {\bf{F}} {\bf{F}}^{\sf H} {\bf{a}}_{\rm t}(\theta) 
    \\
    &= P \cdot {\text{vec}}\left( {\bf{a}}_{\rm t}^{\sf H}(\theta) {\bf{F}} {\bf{F}}^{\sf H} {\bf{a}}_{\rm t}(\theta) \right) 
    \\
    &\mathop{=}^{(a)} P \cdot \left(\left({\bf{a}}_{\rm t}^{\sf T}(\theta){\bf{F}}^* \right) \otimes {\bf{a}}_{\rm t}^{\sf H}(\theta) \right) {\text{vec}}({\bf{F}}) 
    \\\nonumber
    &\mathop{=}^{(b)} P \cdot \left(\left(\left({\bf{I}} \otimes {\bf{a}}_{\rm t}^{\sf T}(\theta) \right) {\text{vec}}({\bf{F}}^*) \right)^{\sf T}  \otimes {\bf{a}}_{\rm t}^{\sf H}(\theta)\right) {\text{vec}}({\bf{F}}) 
    \\
    &\mathop{=}^{} \bar {\bf{f}}^{\sf H} \left(P \cdot {\bf{I}} \otimes {\bf{a}}_{\rm t}(\theta) \otimes {\bf{a}}_{\rm t}^{\sf H}(\theta) \right) \bar {\bf{f}},
\end{align}
where $(a)$ and $(b)$ follow from ${\text {vec}}({\bf{ABC}}) = ({\bf{C}}^{\sf T}\otimes {\bf{A}}){\text {vec}}({\bf{B}})$.   
Let us define ${\bf{A}}(\theta) = P \cdot {\bf{I}} \otimes {\bf{a}}_{\rm t}(\theta) \otimes {\bf{a}}_{\rm t}^{\sf H}(\theta)$. 
Then, the MSE of the radar beam pattern becomes 
\begin{align} \label{eq:mse_reform_1}
    {\text{MSE}}_{\text R}({\bf F}) = \frac{1}{L} \sum_{\ell = 1}^{L} |P d(\theta_\ell) - \bar {\bf{f}}^{\sf H} {\bf{A}}(\theta_{\ell}) \bar {\bf{f}} |^2.
\end{align}
Assuming full power transmission, i.e., $\| \bar {\bf{f}} \|^2 = 1$, \eqref{eq:mse_reform_1} is further rewritten as
\begin{align}
       \label{eq:mse_reform_2}
    {\text{MSE}}_{\text R}({\bf F}) 
    = \frac{1}{L} {\sum_{\ell = 1}^{L} |\bar {\bf{f}}^{\sf H} {\bf{D}}(\theta_\ell) \bar {\bf{f}} - \bar {\bf{f}}^{\sf H} {\bf{A}}(\theta_{\ell}) \bar {\bf{f}} |^2}, 
\end{align}
where ${\bf{D}}(\theta_{\ell}) = P\cdot d(\theta_\ell) \cdot {\bf{I}}_{N(K+M)}$. 

Lastly, we relax the problem by omitting the power constraint ${\rm Tr}({\bf FF}^{\sf H}) = \|\bar {\bf{f}} \|^2 = 1$ and consider it back in the algorithm to obtain the solution within the feasible set.
Applying \eqref{eq:R_lb_rayleigh} and \eqref{eq:mse_reform_2} to the problem in \eqref{eq:prob_formulation}, our main problem is reformulated as
\begin{align} 
    \label{eq:prob_reformulation_omit_power}
    \mathop{\text {maximize}}_{\bar {\bf{f}}}\;\; & \bar R_\Sigma^{\rm lb}(\bar {\bf f})=\sum_{k=1}^K\bar R_k^{\rm lb}(\bar {\bf f})
    \\\label{eq:const_refomulation_omit_power}
    {\text {subject}\;\text {to}} \;\; & \frac{1}{L} \sum_{\ell = 1}^{L} | \bar {\bf{f}}^{\sf H} {\bf{D}}(\theta_\ell) \bar {\bf{f}} - \bar {\bf{f}}^{\sf H} {\bf{A}}(\theta_\ell) \bar {\bf{f}} |^2 \le  T_{\rm mse}. 
\end{align}
We remark that this reformulated problem is still non-convex. 
Accordingly, it is NP-hard to find a global optimal solution, and instead, we show how to obtain a  good local optimal solution of the reformulated and relaxed problem \eqref{eq:prob_reformulation_omit_power}.

\subsection{Proposed Algorithm with MSE Constraint}
\label{sec:optimality}
To identify a local optimal solution, we study the first-order optimality condition;
we derive the stationarity condition of the KKT condition for  \eqref{eq:prob_reformulation_omit_power} in the following lemma:
\begin{lemma} \label{lem:main}
    The KKT stationarity condition of the problem in \eqref{eq:prob_reformulation_omit_power} is satisfied if the following condition holds:
    \begin{align} \label{eq:lemma_kkt}
     {\bf{\Psi}}(\bar {\bf{f}}) \bar {\bf{f}} = {\bf{\Omega}}(\bar {\bf{f}})\lambda(\bar {\bf{f}}) \bar {\bf{f}},
    \end{align}
    where 
    \begin{align} 
    \label{eq:Psi}
    &{\bf{\Psi}}(\bar {\bf{f}}) =  \lambda_{\sf num}(\bar {\bf{f}}) \left[ \sum_{k = 1}^{K}    \frac{{\bf{B}}_{k}  }{\bar {\bf{f}}^{\sf H} {\bf{B}}_{k} \bar {\bf{f}}} + \right.
    \\\nonumber
    &\left. \qquad  \frac{2 \mu\log2}{LP^2} \sum_{\ell = 1}^{L} \left\{ ( \bar {\bf{f}}^{\sf H} {\bf{D}}(\theta_\ell) \bar {\bf{f}}) {\bf{A}}(\theta_\ell) + (\bar {\bf{f}}^{\sf H} {\bf{A}}(\theta_\ell) \bar {\bf{f}})  {\bf{D}}(\theta_\ell) \right\} \right], 
    \\\label{eq:Omega}
    & {\bf{\Omega}}(\bar {\bf{f}}) = \lambda_{\sf den} (\bar {\bf{f}}) \left[ \sum_{k = 1}^{K}    \frac{{\bf{C}}_{k}  }{\bar {\bf{f}}^{\sf H} {\bf{C}}_{k} \bar {\bf{f}}} + \right.
    \\\nonumber
    &\left. \qquad\frac{2 \mu \log2}{LP^2} \sum_{\ell = 1}^{L} \left\{ ( \bar {\bf{f}}^{\sf H} {\bf{D}}(\theta_\ell) \bar {\bf{f}})  {\bf{D}}(\theta_\ell) + (\bar {\bf{f}}^{\sf H} {\bf{A}}(\theta_\ell) \bar {\bf{f}}) {\bf{A}}(\theta_\ell) \right\} \right]. 
    \end{align}
    Here, $\mu$ in \eqref{eq:Psi} and \eqref{eq:Omega} is the Lagrangian multiplier and 
    \begin{align}
        \lambda(\bar {\bf{f}}) &=  \prod_{k = 1}^{K}\left(\frac{\bar{\bf{f}}^{\sf H}{\bf{B}}_k \bar {\bf{f}}}{\bar{\bf{f}}^{\sf H} {\bf{C}}_k\bar{\bf{f}}} \right)  2^{\mu \left(\frac{T_{\rm mse}}{P^2} - \frac{1}{LP^2} \sum_{\ell = 1}^{L} |  \bar {\bf{f}}^{\sf H} {\bf{D}}(\theta_\ell) \bar {\bf{f}} - \bar {\bf{f}}^{\sf H} {\bf{A}}(\theta_\ell) \bar {\bf{f}} |^2    \right)}
        \\\label{eq:lambda}
        &= \frac{\lambda_{\sf num}(\bar {\bf{f}})}{\lambda_{\sf den}(\bar {\bf{f}})},
    \end{align}
    where $\lambda_{\sf num}(\bar {\bf{f}})$ and $ \lambda_{\sf den}(\bar {\bf{f}})$ are any functions that satisfy \eqref{eq:lambda}.
\end{lemma}
\begin{proof}
See Appendix \ref{append:lem1}. 
\end{proof}

If the problem in \eqref{eq:prob_reformulation_omit_power} is convex, we can reach a global optimal point by jointly finding the Lagrangian multiplier $\mu$ and  $\bar {\bf f}$ by solving the KKT condition including the stationarity condition in \eqref{eq:lemma_kkt}. 
Since our problem is non-convex, however, it is infeasible to find a global optimal point with the derived condition, and only a necessary condition of the local optimal point is guaranteed.
In addition, the Lagrangian multiplier $\mu$ and the optimization variable $\bar {\bf{f}}$ are complicatedly intertwined, making it challenging to jointly optimize $\mu$ and $\bar {\bf{f}}$ even for the local optimal point. 
To resolve this difficulty, we alternately optimize $\mu$ and $\bar {\bf{f}}$ by using Lemma~\ref{lem:main}. 
To this end, we have the following interpretation of the condition in \eqref{eq:lemma_kkt}:

\subsubsection{Interpretation}
\label{sec:interpretation}
We know that if a precoding vector $\bar {\bf{f}}$ satisfies \eqref{eq:lemma_kkt}, then it satisfies the stationarity condition.
Although the condition can be met with such a stationary point $\bar {\bf f}$, this point cannot guarantee an efficient solution since it can be a random local optimal point due to the non-convexity of \eqref{eq:prob_reformulation_omit_power}.
To achieve an efficient solution, we first explain how the objective function $\bar R_{\Sigma}^{\sf lb}(\bar {\bf f})$ in \eqref{eq:prob_reformulation_omit_power} and $\lambda(\bar {\bf f})$ in \eqref{eq:lemma_kkt} are related. 
Let us assume that we have any precoder and Lagrangian multiplier pair $(\bar {\bf f}^\dag, \mu^\dag)$  that   satisfies the MSE constraint \eqref{eq:const_refomulation_omit_power} with equality  and the stationarity condition  $\eqref{eq:lemma_kkt}$, i.e., $(\bar {\bf f}^\dag, \mu^\dag)$ meets the necessary condition of the local maximum point.
In such a case, because of the active MSE constraint, we have $ \log_2\lambda(\bar {\bf f}^\dag) = R_{\Sigma}^{\sf lb}(\bar {\bf f}^\dag)$ which is a local maximum from the stationarity condition.
This suggest that we need to find such $(\bar {\bf f}^\dag, \mu^\dag)$ that maximizes $\lambda(\bar {\bf f}^\dag)$ among  stationary points.

To find a {good} stationary point that maximizes $\lambda(\bar {\bf f}^\dag)$, we see \eqref{eq:lemma_kkt} through lens of functional eigenvalue problems. 
Since  ${\bf \Omega}(\bar{\bf f})$ is invertible, we can represent the condition in \eqref{eq:lemma_kkt} as  
\begin{align}
    \label{eq:EigenProblem}
    {\bf{\Omega}}^{-1}(\bar {\bf{f}}){\bf{\Psi}}(\bar {\bf{f}}) \bar {\bf{f}} = \lambda(\bar {\bf{f}}) \bar {\bf{f}}.
\end{align}
Interpreting  $\bar {\bf{f}}$ as an eigenvector  and $\lambda(\bar {\bf{f}})$ as a corresponding eigenvalue of   ${\bf{\Omega}}^{-1}(\bar {\bf{f}}){\bf{\Psi}}(\bar {\bf{f}})$, the KKT stationarity condition \eqref{eq:lemma_kkt} is equivalent to a functional eigenvalue problem. More rigorously, \eqref{eq:lemma_kkt} is cast as a NEPv \cite{cai:siam:18}.
As described in \cite{cai:siam:18}, NEPv is a generalized version of an eigenvector problem, in that a corresponding matrix is changed depending on an eigenvector in a non-linear fashion.
Recall that we need to find  $(\bar {\bf f}^\dag, \mu^\dag)$ which maximizes $\lambda(\bar {\bf f}^\dag)$ to maximize the objective function among the stationary points.
To this end, we should derive the principal eigenvector of \eqref{eq:EigenProblem}, $\bar {\bf f}^\star$, with corresponding $\mu^\star$ that keeps the MSE constraint active, thereby maximizing $\lambda(\bar {\bf f})$ among the stationary points.
Then such point $\bar {\bf f}^\star$ is likely to be the superior local optimal point.
This observation is summarized in the following proposition:
\begin{proposition} 
    \label{prop:optimal}
    Consider that $\bar {\bf{f}}^{\star}$ is the principal eigenvector of ${\bf{\Omega}}^{-1} (\bar {\bf{f}}^{\star}) {\bf{\Psi}}(\bar {\bf{f}}^{\star})$ satisfying
    \begin{align}
        \label{eq:principal}
        {\bf{\Omega}}^{-1}(\bar {\bf{f}}^\star){\bf{\Psi}}(\bar {\bf{f}}^\star) \bar {\bf{f}}^\star = \lambda(\bar {\bf f}^{\star}) \bar {\bf{f}}^\star
    \end{align}
    with corresponding $\mu^\star$ where $(\bar {\bf f}^\star, \mu^\star)$ has an active MSE constraint for \eqref{eq:const_refomulation_omit_power}.
    Then $\bar {\bf f}^\star$ is the superior stationary point for the problem in \eqref{eq:prob_reformulation_omit_power}.
\end{proposition}

\subsubsection{GPI-ISAC Algorithm}
Based on Proposition~\ref{prop:optimal}, we develop an efficient algorithm to identify the principal eigenvector, named as generalized power iteration-based ISAC (GPI-ISAC). 
The algorithm alternately optimizes $\bar {\bf f}$ for given $\mu$ and $\mu$ for given $\bar {\bf f}$ until it finds $(\bar {\bf f}^\star, \mu^\star)$ which is the point shown in Proposition~\ref{prop:optimal}.
In other words, GPI-ISAC seeks for the superior stationary point.
We first propose the ISAC beamforming optimization method for given $\mu$ which iteratively updates $\bar {\bf f}$ \cite{choi:twc:20}.
At the $t$th iteration,  we construct the matrices ${\bf{\Psi}}(\bar {\bf{f}}_{(t-1)})$ and ${\bf{\Omega}}(\bar {\bf{f}}_{(t-1)})$ using \eqref{eq:Psi} and \eqref{eq:Omega}, respectively using $\bar {\bf{f}}_{(t-1)}$ obtained at the $(t-1)$th iteration. Then, we update $\bar {\bf{f}}_{(t)}$ as 
\begin{align}
    \bar {\bf{f}}_{(t)} \leftarrow \frac{ {\bf{\Omega}}^{-1} (\bar {\bf{f}}_{(t-1)}) {\bf{\Psi}} (\bar {\bf{f}}_{(t-1)}) \bar {\bf{f}}_{(t-1)}}{\| {\bf{\Omega}}^{-1} (\bar {\bf{f}}_{(t-1)})  {\bf{\Psi}} (\bar {\bf{f}}_{(t-1)}) \bar {\bf{f}}_{(t-1)} \|}.
\end{align}
We repeat this process until the convergence criterion is met. 
For instance,  $\left\|\bar {\bf{f}}_{(t)} - \bar {\bf{f}}_{(t-1)} \right\| < \epsilon$ for $\epsilon >0$ can be adopted as the criterion. 
Once converges, the obtained  $\bar {\bf{f}}$ is the principal eigenvector of ${\bf{\Omega}}^{-1} (\bar {\bf{f}}) {\bf{\Psi}} (\bar {\bf{f}})$.
Algorithm \ref{alg:gpi_beamforming} summarizes the steps. 

\begin{algorithm} [t]
\caption{GPI-based beamforming} \label{alg:gpi_beamforming} 
{\bf{initialize}}: $\bar {\bf{f}}_{(0)}$\\
Set the iteration count $t = 1$.\\
\While {$\left\|\bar {\bf{f}}_{(t)} - \bar {\bf{f}}_{(t-1)} \right\| > \epsilon$ or $t < t_{\rm max}$ }
{
Build matrix $  {\bf{\Psi}}(\bar {\bf{f}}_{(t-1)})$ according to  \eqref{eq:Psi}.\\
Build matrix $ {\bf{\Omega}} (\bar {\bf{f}}_{(t-1)})$ according to  \eqref{eq:Omega}. \\
Compute $\bar {\bf{f}}_{(t)} \leftarrow \frac{ {\bf{\Omega}}^{-1} (\bar {\bf{f}}_{(t-1)}) {\bf{\Psi}} (\bar {\bf{f}}_{(t-1)}) \bar {\bf{f}}_{(t-1)}}{\|  {\bf{\Omega}}^{-1}  (\bar {\bf{f}}_{(t-1)}) {\bf{\Psi}}(\bar {\bf{f}}_{(t-1)}) \bar {\bf{f}}_{(t-1)} \|}$.
\\
$t \leftarrow t+1$.}
$\bar {\bf{f}} \leftarrow \bar {\bf{f}}_{(t)}$\\
\Return{\ }{$\bar {\bf{f}}$}.
\end{algorithm}

\begin{remark} \normalfont (Complexity of Algorithm~\ref{alg:gpi_beamforming}) 
\label{rm:complexity}
The complexity of the power iteration in the GPI-based beamforming algorithm is dominated by the inversion of $ {\bf{\Omega}}(\bar {\bf{f}})$ in line 6 in Algorithm~\ref{alg:gpi_beamforming}.
As the matrix ${\bf{\Omega}}(\bar {\bf{f}})$ is the sum of $(K+M)$ block diagonal matrices, we can obtain the $ {\bf{\Omega}}^{-1}(\bar {\bf{f}})$ by calculating the inverse of each submatrix of size $N\times N$ instead of computing the inverse of the entire matrix ${\bf \Omega}(\bar {\bf f})$ of size $N(K+M)\times N(K+M)$.
Consequently, the inversion requires the complexity with the order of $\CMcal{O}(N^3(K+M))$.
Based on the analysis, the computational complexity of Algorithm~\ref{alg:gpi_beamforming} is $\CMcal{O}(\xi_{\rm gpi}N^3(K+M))$ where $\xi_{\rm gpi}$ is the number of iterations.
\end{remark}


 Now we explain how to update the proper Lagrangian multiplier $\mu$ for given $\bar {\bf f}$. 
In  the Lagrangian function \eqref{eq:lagrangian}, the Lagrangian multiplier $\mu$ strikes a balance between maximizing the objective function and satisfying the constraint by minimizing the radar beam pattern MSE.
On the one hand, if $\mu$ is large, the optimization process more pursues to minimize the radar beam pattern MSE, so as to satisfy the constraint. In this case, the objective function, i.e., the sum SE, is not efficiently maximized. On the other hand, if $\mu$ is small, the optimization process strives to increase the objective function, while less caring to meet the constraint. 
For this reason, $\mu$ should be set to a proper value to efficiently maximize the objective function with satisfying the MSE constraint.


To identify proper $\mu$, we present a binary tree search based method. 
We assume that the minimum and maximum values of $\mu$ are given, i.e., $\mu_{\min}$ and $\mu_{\max}$. Since the Lagrangian multiplier is non-negative, we set $\mu_{\min} = 0$. 
Once we obtain a stationary point $\bar {\bf{f}}$ for given $\mu$, we check if $T_{\rm mse} > \frac{1}{L} \sum_{\ell = 1}^{L} |  P d(\theta_\ell) - B({\bf F};\theta_{\ell}) |^2$. 
If this is true, we use smaller $\mu$ to further identify the stationary point with less MSE penalty, i.e., more weight on the objective function. 
Accordingly, we decrease $\mu$ by $\mu \leftarrow (\mu + \mu_{\min})/2$ for such a case. 
On the contrary, if $T_{\rm mse} <  \frac{1}{L} \sum_{\ell = 1}^{L} |  P d(\theta_\ell) - B({\bf F};\theta_{\ell}) |^2$, we increase $\mu$ as $\mu \leftarrow (\mu + \mu_{\max})/2$. 
Then we perform Algorithm~\ref{alg:gpi_beamforming} with the updated $\mu$.
Thanks to the binary search, the search space becomes half of the previous search space as the algorithm progresses.
We repeat this process until a convergence condition is satisfied, e.g., $|\mu_{(n)} - \mu_{(n-1)}|<\zeta$ for $\zeta > 0$.
When the problem in \eqref{eq:prob_reformulation_omit_power} is feasible, we can find the smallest $\mu$ that guarantees the MSE requirement, and attain the solution ($\bar {\bf f}^\star, \mu^\star$) in Proposition~\ref{prop:optimal} accordingly. 
We describe the entire optimization steps in Algorithm \ref{alg:gpi_isac}.
\begin{algorithm} [t]
\caption{GPI-ISAC} \label{alg:gpi_isac} 
{\bf{initialize}}: $\mu_{(1)} = \mu_{\max}$.\\ 
Set the iteration count $n= 1$ and \\
\While {$|\mu_{(n)} - \mu_{(n-1)}| > \zeta $ or $n \leq n_{\max} $}
{
$\bar {\bf f} \leftarrow$ Algorithm \ref{alg:gpi_beamforming} with $\mu_{(n)}$\\
\If {$T_{\rm mse} > \frac{1}{L} \sum_{\ell = 1}^{L} |  P d(\theta_\ell) - B({\bf F};\theta_{\ell}) |^2$}
{
$\mu_{(n+1)} \leftarrow (\mu_{(n)} + \mu_{\min})/2$
}
\Else 
{
$\mu_{(n+1)} \leftarrow (\mu_{(n)} + \mu_{\max})/2$
}
$n \leftarrow n+1$.
}
$\bar {\bf f}^\star \leftarrow \bar {\bf f}$\\
\Return {$\bar{\bf{f}}^{\star}$}
\end{algorithm}

\begin{remark} \normalfont (Complexity of Algorithm~\ref{alg:gpi_isac}) \label{rm:complexity_total}
In Algorithm~\ref{alg:gpi_isac}, the required complexity for performing the feasibility check in line 5 needs $\CMcal{O}(LN(K+M))$. 
Since the complexity of Algorithm~\ref{alg:gpi_beamforming} is $\CMcal{O}(\xi_{\rm gpi}N^3(K+M))$, the total computational complexity of Algorithm~\ref{alg:gpi_isac} is 
$\CMcal{O}(\xi_{\mu}(LN(K+M) + \xi_{\rm gpi}N^3(K+M)))$.
When $L \leq  \xi_{\rm gpi}N^2$, $\CMcal{O}(\xi_{\mu}\xi_{\rm gpi}  N^3(K+M))$ is the  complexity of the GPI-ISAC algorithm which can be considered to be general.
\end{remark}





\subsection{Proposed Algorithm with SCNR Constraint}

Now, the similar framework is applied for solving the SCNR constrained problem in \eqref{eq:prob_formulation_scnr}; the SCNR in \eqref{eq:scnr} is rewritten as
\begin{align}
   \gamma_{\rm R} & = \frac{{\rm vec}({\bf G}_{\rm tar}{\bf F})^{\sf H}{\rm vec}({\bf G}_{\rm tar}{\bf F})}{{\rm vec}({\bf G}_{\rm cl}{\bf F})^{\sf H}{\rm vec}({\bf G}_{\rm cl}{\bf F}) + \frac{N}{P}\sigma^2_{\rm R}}
   \\
    & \stackrel{(a)}= \frac{\bar {\bf f}^{\sf H}({\bf I}_{K+M}\otimes{\bf G}_{\rm tar})^{\sf H}({\bf I}_{K+M}\otimes{\bf G}_{\rm tar})\bar {\bf f}}{\bar {\bf f}^{\sf H}({\bf I}_{K+M}\otimes{\bf G}_{\rm cl})^{\sf H}({\bf I}_{K+M}\otimes{\bf G}_{\rm cl})\bar {\bf f}+ \frac{N}{P}\sigma^2_{\rm R}},
\end{align}
where $(a)$ comes from ${\text {vec}}({\bf{ABC}}) = ({\bf{C}}^{\sf T}\otimes {\bf{A}}){\text {vec}}({\bf{B}})$.
Similarly to the MSE-constrained problem in \eqref{eq:prob_reformulation_omit_power}, we assume $\|\bar {\bf f}\|^2=1$ and replace $R_{\Sigma}({\bf F})$ in \eqref{eq:prob_formulation_scnr} with  $\bar R_{\Sigma}^{\sf lb}(\bar {\bf f})$ in \eqref{eq:R_lb_rayleigh}.
Then the SCNR-constrained problem in \eqref{eq:prob_formulation_scnr} is  reformulated as
\begin{align} 
    \label{eq:prob_reformulation_scnr}
    \mathop{\text {maximize}}_{\bar {\bf{f}}}\;\; & \bar R_\Sigma^{\rm lb}(\bar {\bf f})
    \\
    {\text {subject}\;\text {to}} \;\; & \frac{\bar {\bf f}^{\sf H} {\bar {\bf G}}_{\rm tar}\bar {\bf f}}{\bar {\bf f}^{\sf H}{\bar {\bf G}}_{\rm cl}\bar {\bf f}} \ge  T_{\rm scnr},
\end{align}
where 
\begin{align}
    \label{eq:barGtar}
    &\bar {\bf G}_{\rm tar} =({\bf I}_{K+M}\otimes{\bf G}_{\rm tar})^{\sf H}({\bf I}_{K+M}\otimes{\bf G}_{\rm tar}),
    \\\label{eq:barGcl}
    &\bar {\bf G}_{\rm cl} = ({\bf I}_{K+M}\otimes{\bf G}_{\rm cl})^{\sf H}({\bf I}_{K+M}\otimes{\bf G}_{\rm cl})+ \frac{N}{P}\sigma^2_{\rm R}{\bf{I}}_{N(K+M)}.
\end{align}
The power constraint $\|\bar {\bf f}\|^2 = 1$ is naturally omitted since the problem in \eqref{eq:prob_reformulation_scnr} is invariant up to scaling of  $\bar {\bf f}$.

Next, we also derive the KKT stationarity condition of the reformulated SCNR-constrained problem in \eqref{eq:prob_reformulation_scnr}.
\begin{lemma} 
    \label{lem:main2}
    The KKT stationarity condition of the problem in \eqref{eq:prob_reformulation_scnr} is satisfied if the following condition holds
    \begin{align} 
        \label{eq:lemma_kkt_scnr}
        {\bf{\Upsilon}}(\bar {\bf{f}}) \bar {\bf{f}} = {\bf{\Xi}}(\bar {\bf{f}})\eta(\bar {\bf{f}}) \bar {\bf{f}},
    \end{align}
    where 
    \begin{align} 
    \label{eq:Upsilon}
    &  {\bf{\Upsilon}}(\bar {\bf{f}}) =   \eta_{\sf num}(\bar {\bf f})\left( \sum_{k = 1}^{K}    \frac{{\bf{B}}_{k}  }{\bar {\bf{f}}^{\sf H} {\bf{B}}_{k} \bar {\bf{f}}} + \mu  \frac{ {\bar {\bf G}}_{\rm tar}}{\bar {\bf f}^{\sf H}{\bar {\bf G}}_{\rm tar}\bar {\bf{f}} }   \right) , 
    \\\label{eq:Phi}
    & {\bf{\Xi}}(\bar {\bf{f}}) = \eta_{\sf den}(\bar {\bf f})\left( \sum_{k = 1}^{K}    \frac{{\bf{C}}_{k}  }{\bar {\bf{f}}^{\sf H} {\bf{C}}_{k} \bar {\bf{f}}} + \mu  \frac{ {\bar {\bf G}}_{\rm cl}}{\bar {\bf f}^{\sf H}{\bar {\bf G}}_{\rm cl}\bar {\bf{f}} } \right). 
    \end{align}
    Here, $\mu$ in \eqref{eq:Upsilon} and \eqref{eq:Phi} is the Lagrangian multiplier and 
    \begin{align}
        \label{eq:eta}
        \eta(\bar {\bf{f}}) &=  {T^{-\mu}_{\rm scnr}}\left(\frac{\bar {\bf f}^{\sf H} {\bar {\bf G}}_{\rm tar}\bar {\bf f}}{\bar {\bf f}^{\sf H}{\bar {\bf G}}_{\rm cl}\bar {\bf f}}\right)^\mu \prod_{k = 1}^{K}\left(\frac{\bar{\bf{f}}^{\sf H}{\bf{B}}_k \bar {\bf{f}}}{\bar{\bf{f}}^{\sf H} {\bf{C}}_k\bar{\bf{f}}} \right) = \frac{\eta_{\sf num}(\bar {\bf{f}})}{\eta_{\sf den}(\bar {\bf{f}})}.
    \end{align}
    where $\eta_{\sf num}(\bar {\bf{f}})$ and $ \eta_{\sf den}(\bar {\bf{f}})$ are any functions that satisfy \eqref{eq:eta}.
\end{lemma}
\begin{proof}
See Appendix \ref{append:lem2}. 
\end{proof}
Similar to the MSE case, we have the following interpretation as follows: 
\begin{proposition} 
    \label{prop:optimal_scnr}
    Consider that $\bar {\bf{f}}^{\star}$ is the principal eigenvector of ${\bf{\Xi}}^{-1} (\bar {\bf{f}}^{\star}) {\bf{\Upsilon}}(\bar {\bf{f}}^{\star})$ satisfying ${\bf{\Xi}}^{-1}(\bar {\bf{f}}^\star){\bf{\Upsilon}}(\bar {\bf{f}}^\star) \bar {\bf{f}}^\star = \eta(\bar {\bf f}^{\star}) \bar {\bf{f}}^\star$
    with corresponding $\mu^\star$ where $(\bar {\bf f}^\star, \mu^\star)$ has an active SCNR constraint.
    Then $\bar {\bf f}^\star$ is the superior stationary point for \eqref{eq:prob_reformulation_scnr}.
\end{proposition}
Based on Lemma~\ref{lem:main2} and Proposition~\ref{prop:optimal_scnr}, the GPI-ISAC algorithm that solves the SCNR-constrained problem in \eqref{eq:prob_reformulation_scnr} can be developed by replacing ${\bf \Psi}(\bar {\bf f}_{(t-1)})$ and ${\bf \Omega}(\bar {\bf f}_{(t-1)})$  in Algorithm~\ref{alg:gpi_beamforming} with ${\bf \Upsilon}(\bar {\bf f}_{(t-1)})$ and ${\bf \Xi}(\bar {\bf f}_{(t-1)})$, respectively, and the feasibility check in line 5 of Algorithm~\ref{alg:gpi_isac} with $ T_{\rm scnr} \leq \frac{\bar {\bf f}^{\sf H} {\bar {\bf G}}_{\rm tar}\bar {\bf f}}{\bar {\bf f}^{\sf H}{\bar {\bf G}}_{\rm cl}\bar {\bf f}}$. 
Consequently, the complexity of GPI-ISAC for the SCNR-constrained problem remains the same as  $\CMcal{O}(\xi_{\mu}\xi_{\rm gpi} N^3(K+M))$.


\begin{remark} 
    \normalfont (Complexity Comparison) \label{rm:complexity_comparison}
    The radar-oriented ISAC beamforming method was proposed in \cite{liu:tsp:20} by using a semidefinite relaxation (SDR) method.
    Assuming that $M = N$, the SDR-based method has the complexity of $\CMcal{O}(K^{6.5} N^{6.5} \log (1/\epsilon))$ associated with solving quadratic semidefinite programming (QSDP) with a solution accuracy $\epsilon$. 
    This complexity is higher than that of the proposed method which is $\CMcal{O}(\xi_{\mu}\xi_{\rm gpi} N^3(K+M))$ with the assumption of $L \leq  \xi_{\rm gpi}N^2$.
    In addition, the communication-oriented  weighted minimum mean square error-based majorization minimization (WMMSE-MM) approach proposed in \cite{xu:access:20} has the complexity order of $\CMcal{O}(\xi_{\rm out}\xi_{\rm in}(NK)^3)$ with fixed $\mu$,
    where $\xi_{\rm out}$ and $\xi_{\rm in}$ are the number of iterations for the outer and inner loops of the WMMSE-MM method, respectively. 
    When further considering the same $\mu$ update and the power iteration method for computing principal eigenvalue for the WMMSE-MM, the complexity becomes $\CMcal{O}(\xi_{\mu}\xi_{\rm out}\xi_{\rm in}N^2K)$. 
    We note that the WMMSE-MM approach does not consider a dedicated radar waveform symbol ${\bf s}_{\rm R}$. 
    For fair comparison, the complexity of the proposed GPI-ISAC method without the radar waverform 
    symbols is $\CMcal{O}(\xi_{\mu}\xi_{\rm gpi} N^3K)$.
    Therefore, considering that the magnitudes of the iteration counts and the number of the BS antennas $N$ are comparable, the proposed GPI-ISAC method achieves a similar complexity to the WMMSE-MM approach with the weight optimization.
    In Section~\ref{sec:sim}, we demonstrate that the proposed method outperforms both the SDR and WMMSE-MM methods in the communication and radar performance with lower or similar complexity.
\end{remark}

\section{Extension to Radar-Only Setup}

In this section, we present that the proposed optimization framework can be applicable in the radar-only setup.
A main target becomes minimizing the radar beam pattern MSE in \eqref{eq:MSE} or maximizing the SCNR in \eqref{eq:scnr}.
Firstly, the MSE minimization problem is formulated as follows:
\begin{align} 
    \label{eq:prob_radaronly}
    \mathop{\text {minimize}}_{\bar {\bf{f}}}\;\; &   \sum_{\ell = 1}^{L} |P d(\theta_\ell) - B({\bf F};\theta_{\ell}) |^2
    \\\label{eq:power_radar} 
    {\text {subject}\;\text {to}} \;\; & \| \bar {\bf{f}} \|^2 \leq 1.
\end{align}
Following the same framework, we present the corollary that shows the stationarity condition of \eqref{eq:prob_radaronly}. 
\begin{corollary}
    \label{col:rad_mse}
    The stationarity condition of  \eqref{eq:prob_radaronly}  is derived as
    \begin{align} \label{eq:lemma_kkt_radar}
        {\bf{\Omega}}^{-1}_{\rm R} (\bar {\bf{f}}) {\bf{\Psi}}_{\rm R} (\bar {\bf{f}}) \bar {\bf{f}} = \lambda_{\rm R}(\bar {\bf{f}}) \bar {\bf{f}},
    \end{align}
    where 
    \begin{align} 
        \nonumber
        &  {\bf{\Psi}}_{\rm R}(\bar {\bf{f}}) =  \lambda_{\rm R,\sf num}(\bar {\bf{f}})   \sum_{\ell = 1}^{L} \left\{ ( \bar {\bf{f}}^{\sf H} {\bf{D}}(\theta_\ell) \bar {\bf{f}}) {\bf{A}}(\theta_\ell) + (\bar {\bf{f}}^{\sf H} {\bf{A}}(\theta_\ell) \bar {\bf{f}})  {\bf{D}}(\theta_\ell) \right\}, 
        \\\nonumber
        & {\bf{\Omega}}_{\rm R}(\bar {\bf{f}}) = \lambda_{\rm R,\sf den} (\bar {\bf{f}})   \sum_{\ell = 1}^{L} \left\{ ( \bar {\bf{f}}^{\sf H} {\bf{D}}(\theta_\ell) \bar {\bf{f}})  {\bf{D}}(\theta_\ell) + (\bar {\bf{f}}^{\sf H} {\bf{A}}(\theta_\ell) \bar {\bf{f}}) {\bf{A}}(\theta_\ell) \right\}.
    \end{align}
    and $\lambda_{\rm R}(\bar {\bf{f}}) =    2^{-\sum_{\ell = 1}^{L} |  \bar {\bf{f}}^{\sf H} {\bf{D}}(\theta_\ell) \bar {\bf{f}} - \bar {\bf{f}}^{\sf H} {\bf{A}}(\theta_\ell) \bar {\bf{f}} |^2   }= \frac{\lambda_{\rm R, \sf num}(\bar {\bf{f}})}{\lambda_{\rm R,\sf den}(\bar {\bf{f}})}$.
\end{corollary}
\begin{proof}
Assuming $\|\bar {\bf f}\|^2=1$, we reformulate and convert the minimization to maximization of $-\sum_{\ell = 1}^{L} | \bar {\bf{f}}^{\sf H} {\bf{D}}(\theta_\ell) \bar {\bf{f}} - \bar {\bf{f}}^{\sf H} {\bf{A}}(\theta_\ell) \bar {\bf{f}} |^2 $.
Now the power constraint   \eqref{eq:power_radar} can be ignored.
Then, the rest of the proof is straightforwardly derived from the proof of Lemma~\ref{lem:main} by ignoring the SE term and the MSE threshold with proper simplification. 
\end{proof}
From Corollary~\ref{col:rad_mse}, a GPI-based radar beamforming algorithm can be proposed by adopting Algorithm~\ref{alg:gpi_beamforming} with using $ {\bf{\Psi}}_{\rm R}(\bar {\bf{f}})$  and ${\bf{\Omega}}_{\rm R}(\bar {\bf{f}})$ instead of $ {\bf{\Psi}}(\bar {\bf{f}})$  and ${\bf{\Omega}}(\bar {\bf{f}})$, respectively, to seek for the principal eigenvector of \eqref{eq:lemma_kkt_radar}. 
We remark that the achieved $\bar {\bf f}$ from Algorithm~\ref{alg:gpi_beamforming} with $ {\bf{\Psi}}_{\rm R}(\bar {\bf{f}})$  and ${\bf{\Omega}}_{\rm R}(\bar {\bf{f}})$ is likely to be a superior stationary point of the problem in \eqref{eq:prob_radaronly}.

Now, considering the SCNR maximization, the problem with the SCNR in \eqref{eq:scnr} is formulated  as
\begin{align} 
    \label{eq:prob_radaronly_scnr}
    \mathop{\text {maximize}}_{\bar {\bf{f}}}\;\; &  \gamma_{\rm R} = \frac{{\rm Tr}\left({\bf G}_{\rm tar}{\bf F}{\bf F}^{\sf H}{\bf G}_{\rm tar}^{\sf H}\right)}{{\rm Tr}\left({\bf G}_{\rm cl}{\bf F}{\bf F}^{\sf H}{\bf G}_{\rm cl}^{\sf H}\right) + \frac{N}{P}\sigma^2_{\rm R}}
    \\ 
    {\text {subject}\;\text {to}} \;\; & \| \bar {\bf{f}} \|^2 \leq 1.
\end{align}
\begin{corollary}
    \label{col:rad_scnr}
The optimal solution of the problem \eqref{eq:prob_radaronly_scnr} is the principal eigenvector of the generalized eigenvalue problem:
    \begin{align} 
        \label{eq:lemma_kkt_scnr_radar}
       {\bf{\Xi}}_{\rm R}^{-1} {\bf{\Upsilon}}_{\rm R} \bar {\bf f} = \lambda \bar {\bf f},
    \end{align}
    where $ {\bf{\Upsilon}}_{\rm R} = {\bar {\bf G}}_{\rm tar}$ in \eqref{eq:barGtar} and ${\bf{\Xi}}_{\rm R} =   {\bar {\bf G}}_{\rm cl}$ in  \eqref{eq:barGcl}.
\end{corollary}
\begin{proof}
    Consider $\|\bar {\bf f}\|^2=1$ since it maximizes the SCNR in terms of the transmit power.
    Then following the same reformulation as in Lemma~\ref{lem:main2}, we reformulate the SCNR as
    \begin{align}
       \gamma_{\rm R} = \frac{\bar {\bf f}^{\sf H} {\bar {\bf G}}_{\rm tar}\bar {\bf f}}{\bar {\bf f}^{\sf H}{\bar {\bf G}}_{\rm cl}\bar {\bf f}}.
    \end{align}
    This is a generalized Rayleigh quotient. 
    Since  $\|\bar {\bf f}\|^2=1$ is naturally omitted thank to scaling invariance, the optimal solution is the principal eigenvector of ${\bar {\bf G}}_{\rm cl}^{-1}{\bar {\bf G}}_{\rm tar}$. 
\end{proof}
From Corollary~\ref{col:rad_scnr}, a GPI-based radar beamforming algorithm with the SCNR metric can be proposed by adopting Algorithm~\ref{alg:gpi_beamforming} with using $ {\bf{\Upsilon}}_{\rm R}$  and ${\bf{\Xi}}_{\rm R}$ instead of $ {\bf{\Psi}}(\bar {\bf{f}})$  and ${\bf{\Omega}}(\bar {\bf{f}})$, respectively, to find the principal eigenvector of \eqref{eq:lemma_kkt_scnr_radar}. 
\section{Numerical Results}
\label{sec:sim}

In this section, we evaluate the performance of the proposed methods and draw insights on using different radar performance metrics.
The simulation environment is set as follows: the number of angular sampling grids is set to be $L = 256$ between $-\pi/2$ and $\pi/2$ in radians.
For constructing the channel and its covariance matrix ${\bf K}_k$, we adopt the model presented in \cite{adhi:tit:13}. 
The channel estimation quality parameter is set to be $\kappa = 0.3$.
For algorithms, we consider $\mu_{\rm max} = 2000$ and $\mu_{\rm min} = 0$.
In addition, all convergence thresholds and maximum iteration counts are set to be $10^{-3}$ and $20$, respectively.
We also define the normalized MSE (NMSE) constraint and SNR as $T_{\rm nmse} = T/P^2$ and ${\rm SNR}=P/\sigma^2$, respectively, for brevity in notation. We consider uniform linear array (ULA).
\begin{figure}[t]
    \centering
    \subfigure[Achieved SE]{
    \includegraphics[width=0.9\linewidth]{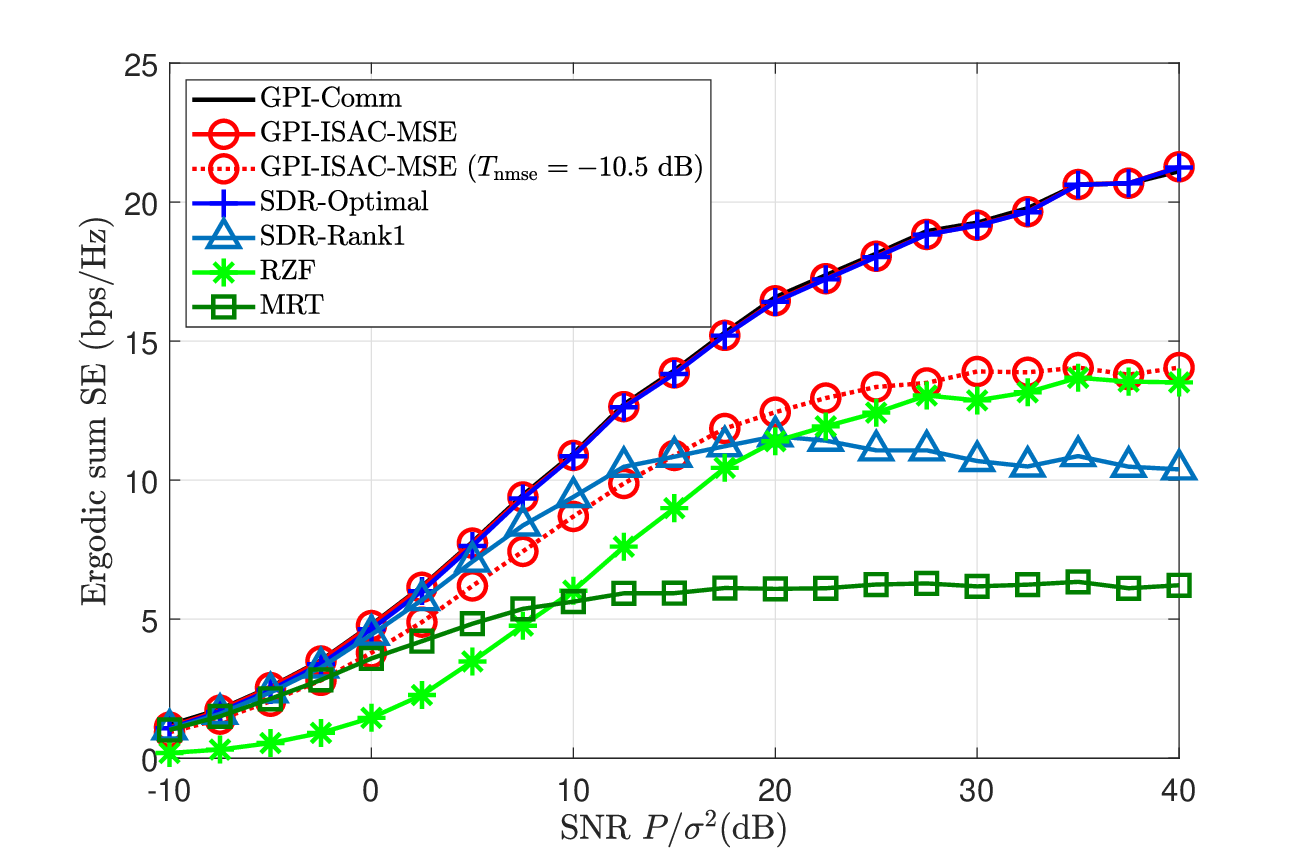}   
    }\\ 
    \centering
    \subfigure[Achieved NMSE]{
    \includegraphics[width=0.9\linewidth]{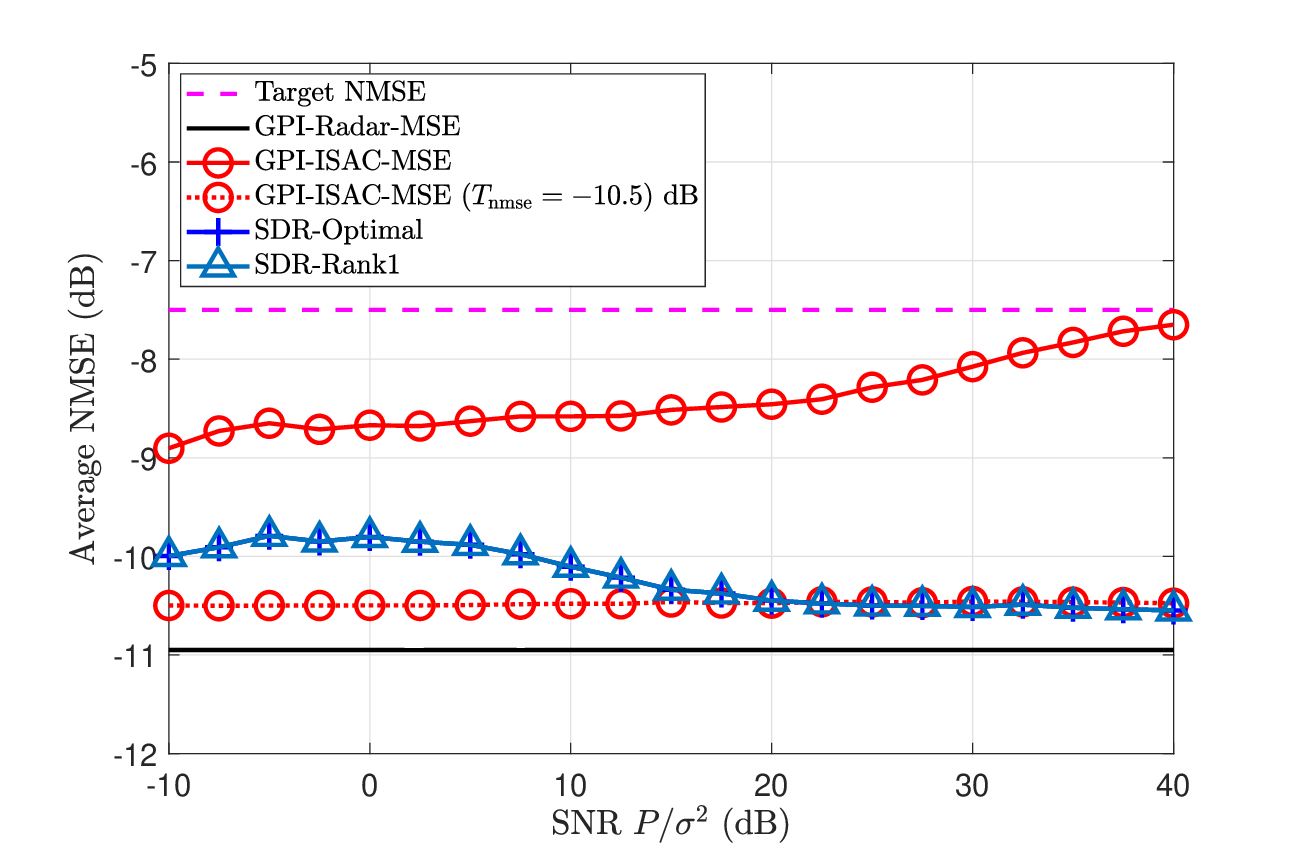}
    }
    \caption{(a) The ergodic sum SE and (b) the average  NMSE with respect to the SNR for $N = 8$, $K = 4$, $M = 8$, and $T_{\rm nmse} = -7.5$ dB. Since SDR-based algorithms achieve a lower NMSE $\approx (-10.5 \sim  -10)$ dB, the GPI-ISAC result with $T_{\rm nmse} = -10.5$ dB is also included for comparison.}
    \label{fig:VsSnr_MSE}
\end{figure}

Now, we briefly introduce the existing approaches to solve optimization problems in ISAC systems for benchmark.

{\bf SDR-based Algorithm:}
Using the proposed optimization method in \cite{liu:twc:18, liu:tsp:20}, the following problem is solved, and beamforming performance is evaluated as a benchmark for GPI-ISAC with the MSE radar metric:
\begin{align} \label{eq:prob_formulation_prior}
    \mathop{\text {minimize}}_{{\bf{F}}}\;\; & \frac{1}{L} \sum_{\ell = 1}^{L} |P d(\theta_\ell) - B({\bf F};\theta_{\ell}) |^2 \\
    {\text {subject}\;\text {to}} \;\; & \gamma_k \ge \Gamma_k,\ \|{\bf{F}} \|_{\sf F} = 1.
\end{align}
Since the radar metric (MSE) is placed in the objective function while the communication metrics (SINR) are placed in the constraints, the SDR-based algorithm can be viewed as a radar-oriented beamforming framework. 
For fair comparison, we set the quality of service (QoS) constraint $\Gamma_k$ as the SINRs achieved from GPI-ISAC (MSE). Since the SDR-based algorithm cannot handle the channel estimation error, it uses an estimated channel to compute  $\gamma_k$. 
Due to the channel estimation error, the problem with such $\Gamma_k$ from GPI-ISAC (MSE) can be infeasible for some cases. 
Then we repeatedly reduce $\Gamma_k,  \forall k$  until the problem becomes feasible. 
We reduce $\Gamma_k$ by $10\%$ at each occurrence.

{\bf WMMSE-MM Algorithm:}
Using the proposed WMMSE-MM method in \cite{xu:access:20}, the following problem is solved:
\begin{align} \label{eq:mmse_mm}
    \mathop{\text {maximize}}_{{\bf{F}}}\;\; &  \sum_{k = 1}^{K} R_k + \mu{\bf{a}}^{\sf H} (\theta^*) {\bf{F}} {\bf{F}}^{\sf H} {\bf{a}}(\theta^*)\\
    {\text {subject}\;\text {to}} \;\;&  \|{\bf{F}} \|_{\sf F} = 1,
\end{align}
where $\theta^*$ is the desired radar beam direction. 
For comparison, we further adopt our binary-search based line search for finding $\mu$ that the derived $\bf F$ satisfies the same SCNR constraint  as our problem in \eqref{eq:prob_formulation_scnr} with, and corresponding beamforming performance is evaluated as a benchmark for GPI-ISAC with the SCNR radar metric for a single-target case.
\subsection{MSE Radar Performance Metric}

\begin{figure} 
    \centering 
    \includegraphics[width=0.85\columnwidth]{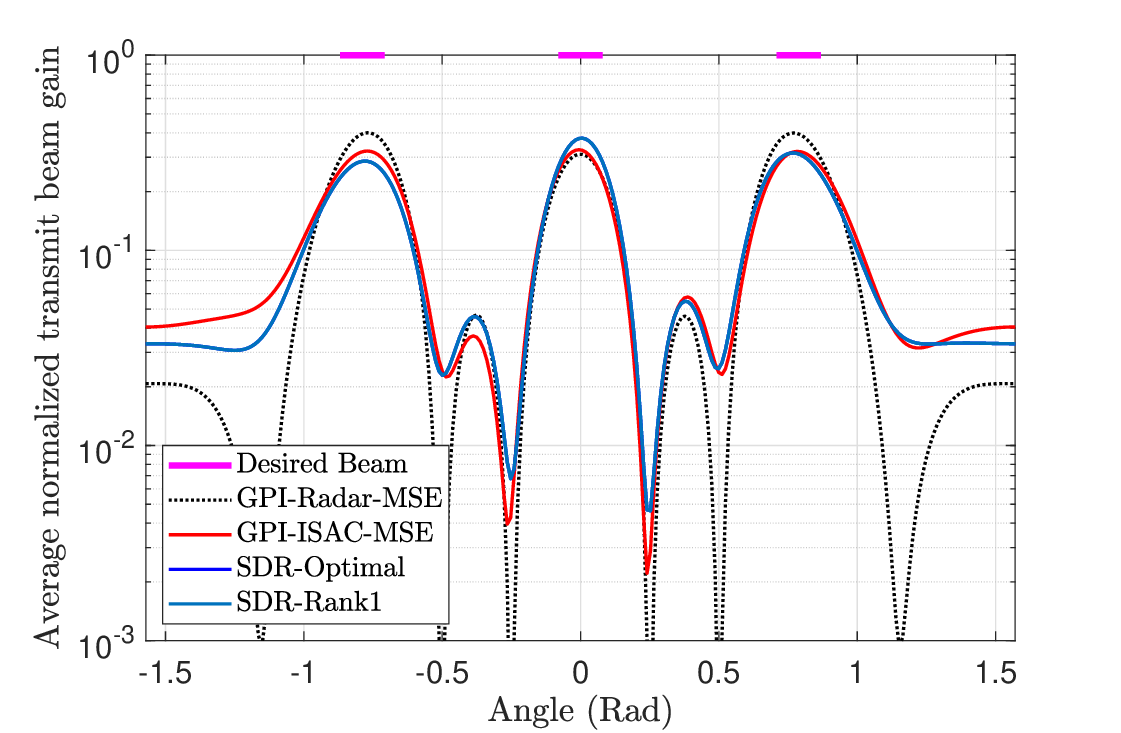}
    \caption{A normalized transmit beam pattern with the achieved NMSE $\approx -10.5$ dB for the GPI-ISAC and SDR-based algorithms at ${\rm SNR} = 25$ dB. Corresponding sum SEs of  GPI-ISAC-MSE and SDR-Rank1 are $13.4$ and $11.1$ ${\rm bps/Hz}$, respectively.}
    \label{fig:TxBeam_MtN8K4}
\end{figure}
We first evaluate GPI-ISAC with the MSE constraint (GPI-ISAC-MSE) for multiple target angles located in $\{-\pi/4, 0, \pi/4\}$ radians with the beam width of $\pi/36$ radian. 
To this end, the desired normalized beam pattern is shaped to be $d(\theta) = 1$, centered around the target angles with the width of $\pi/6$, and $d(\theta) = 0$ elsewhere.
We consider  $N = 8$, $K = 4$, and $M = 8$.
In simulations, GPI-Comm indicates the proposed GPI-ISAC algorithm without the radar  constraint which reduces to the precoding algorithm in \cite{choi:twc:20}. 
Accordingly, it is considered to provide a communication performance upper bound for GPI-ISAC with the MSE radar metric (GPI-ISAC-MSE).
Similarly, GPI-Radar-MSE represents the proposed algorithm for the radar-only setup with radar beam MSE minimization, and hence, its radar MSE performance is regarded as a lower bound for GPI-ISAC-MSE.

\begin{figure} 
    \centering 
    \subfigure[Achieved SE]{\includegraphics[width=0.85\columnwidth]{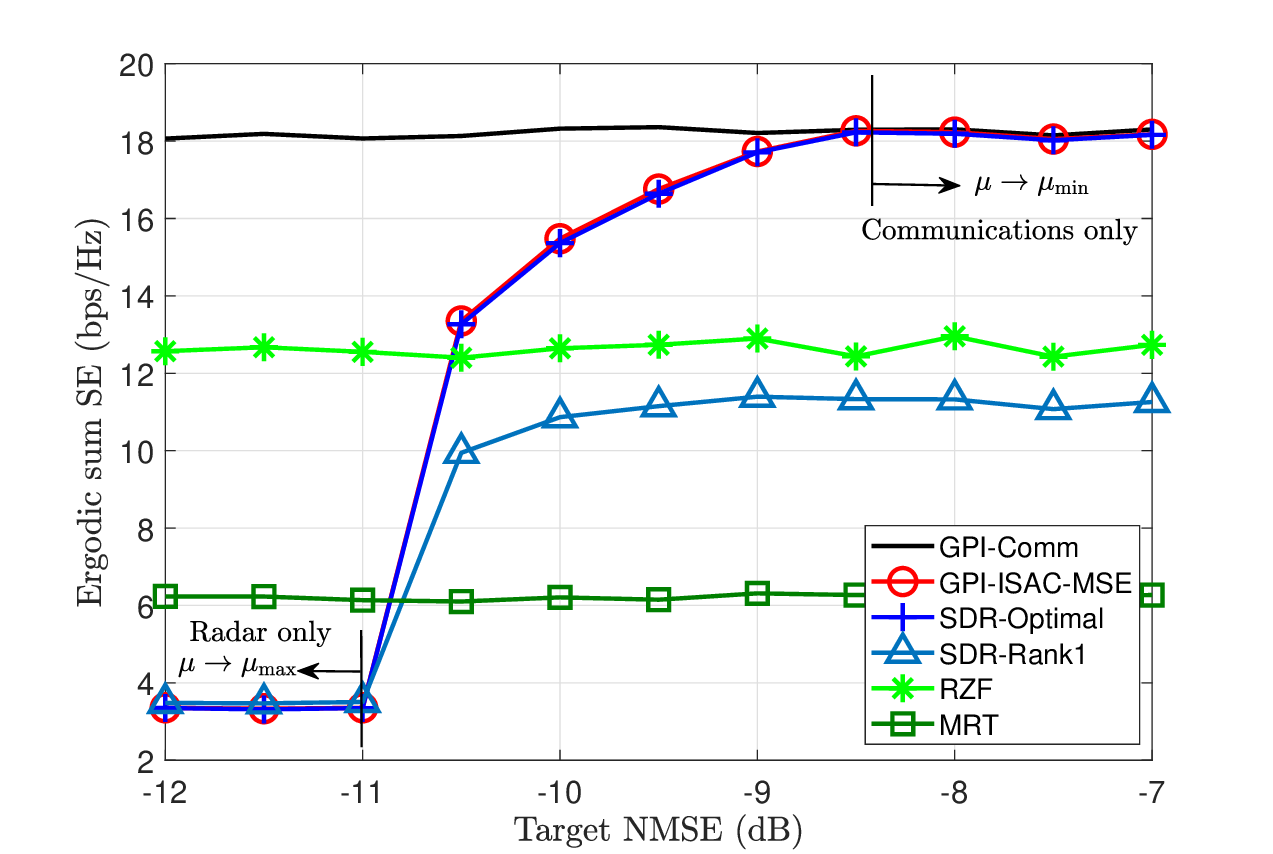}
    }\\
    \centering 
    \subfigure[Achieved NMSE]{\includegraphics[width=0.85\columnwidth]{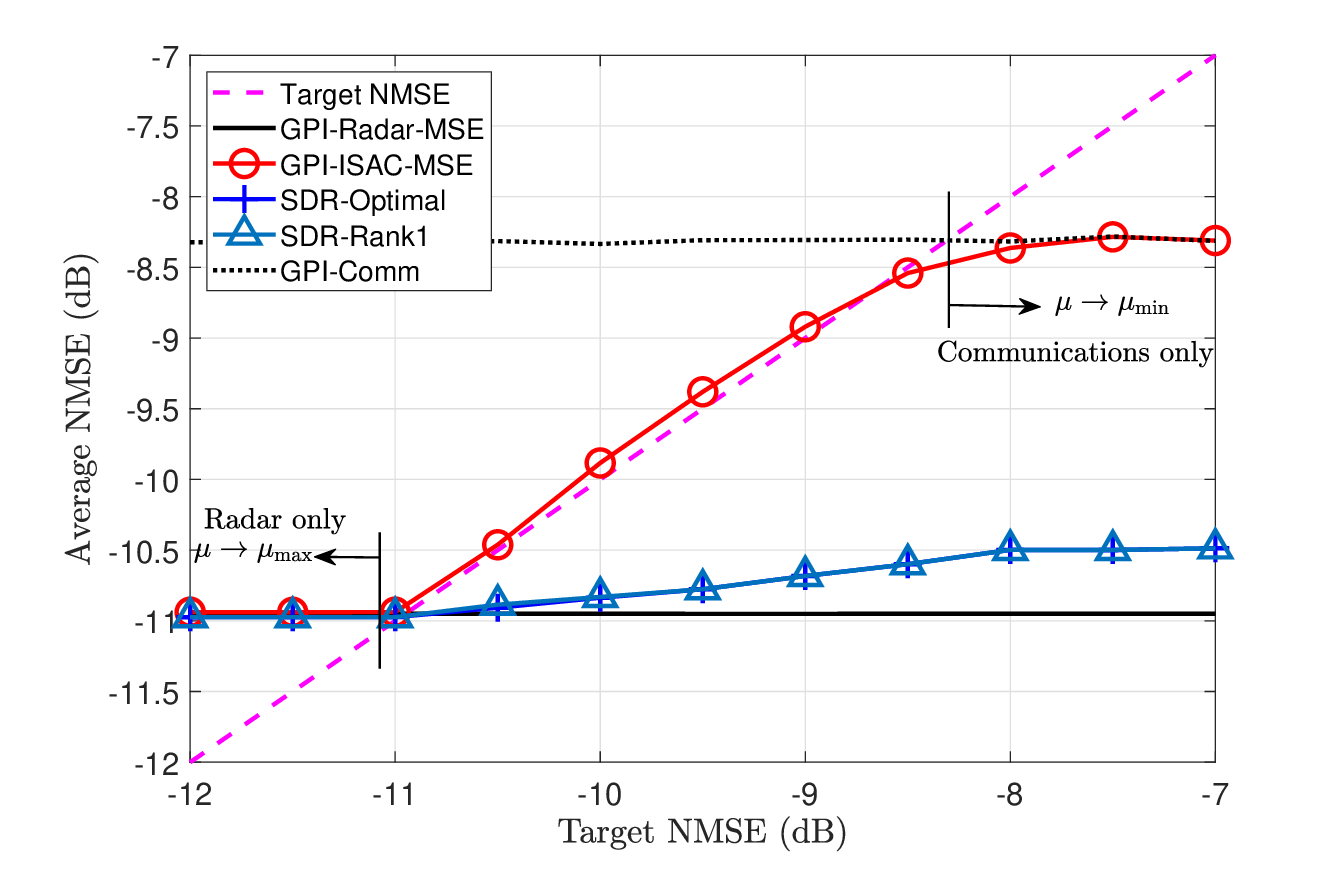}
    }
    \caption{(a) The ergodic sum SE and (b) the average NMSE with respect to the target NMSE $T_{\rm nmse}$ for $N = 8$, $K = 4$, $M = 8$, and ${\rm SNR} = 40$ dB.}
    \label{fig:VsMse_MSE}
\end{figure}

Fig.~\ref{fig:VsSnr_MSE} shows (a) the ergodic sum SE and (b) the average NMSE with respect to the SNR for $T_{\rm nmse} = -7.5$ dB. 
It is observed that the SDR-based algorithm without rank-1 approximation (SDR-Optimal) achieves the same SE as GPI-ISAC-MSE with the lower MSE since it is optimal when ignoring the requirement for the rank-1 approximation. 
Note that this is indeed infeasible to use. 
To obtain the feasible beamforming vectors, rank-1 approximation is requisite. 
Fig.~\ref{fig:VsSnr_MSE} reveals that GPI-ISAC-MSE provides the higher SE than the SDR-based algorithm after rank-1 approximation (SDR-Rank1) and RZF for $T_{\rm nmse} = -7.5$ dB while meeting the MSE constraint.
In this regard, we observe that 
GPI-ISAC-MSE significantly outperforms SDR-Rank1 in terms of the SE while satisfying the MSE constraints.
This is because SDR-Rank1 is a radar-oriented ISAC beamforming algorithm. 
For more comparison, the GPI-ISAC-MSE result with $T_{\rm nmse} = -10.5$ dB is also included in Fig.~\ref{fig:VsSnr_MSE} since the SDR-based algorithms achieve NMSE $\approx -10.5$ dB as shown in Fig.~\ref{fig:VsSnr_MSE}(b).
With the similar achieved NMSE, GPI-ISAC-MSE also offers better SE performance than SDR-Rank1 in most cases.

In Fig.~\ref{fig:TxBeam_MtN8K4}, we further present the corresponding transmit beam patterns of the algorithms evaluated in Fig.~\ref{fig:VsSnr_MSE} for $40$ dB SNR. 
The target beam angles are also marked.
As expected, GPI-Radar-MSE shows the best beam pattern by achieving NMSE $\approx - 11$ dB.
GPI-ISAC-MSE ($T_{\rm nmse} = -10.5$ dB) and SDR-based algorithms provide almost the same beam pattern  with each other as they achieve similar MSE performance while GPI-ISAC-MSE  ($T_{\rm nmse} = -10.5$ dB) has the higher sum SE than SDR-Rank1 by $\sim \! \!3$ bps/Hz. This result implies that GPI-ISAC-MSE is more efficient and effective in providing th higher sum SE while accomplishing the required radar beamforming performance compared to SDR-Rank1.

\begin{figure} 
    \centering 
    \subfigure[Achieved SE]{\includegraphics[width=0.87\columnwidth]{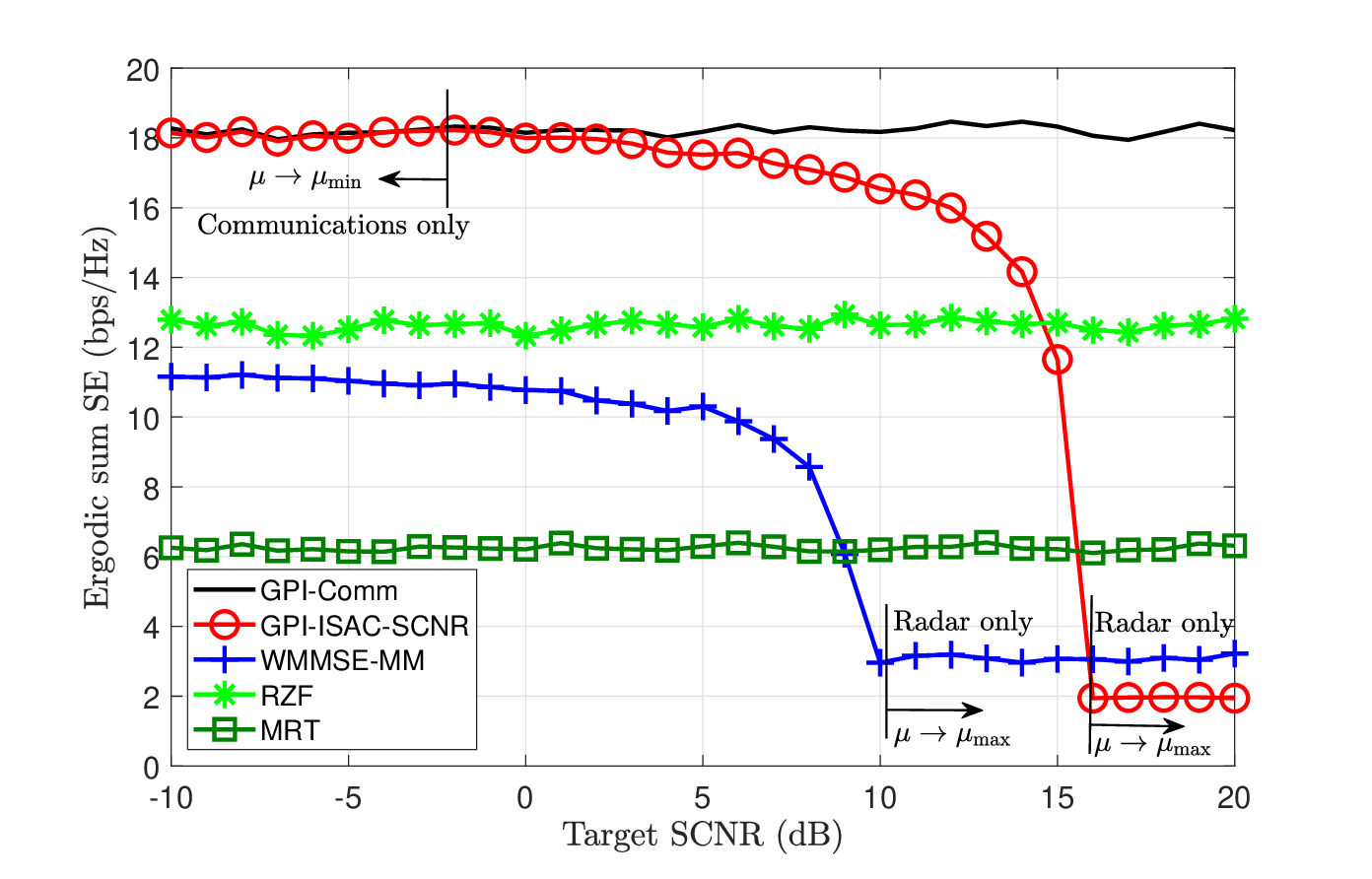}
    }\\
    \centering 
    \subfigure[Achieved SCNR]{\includegraphics[width=0.87\columnwidth]{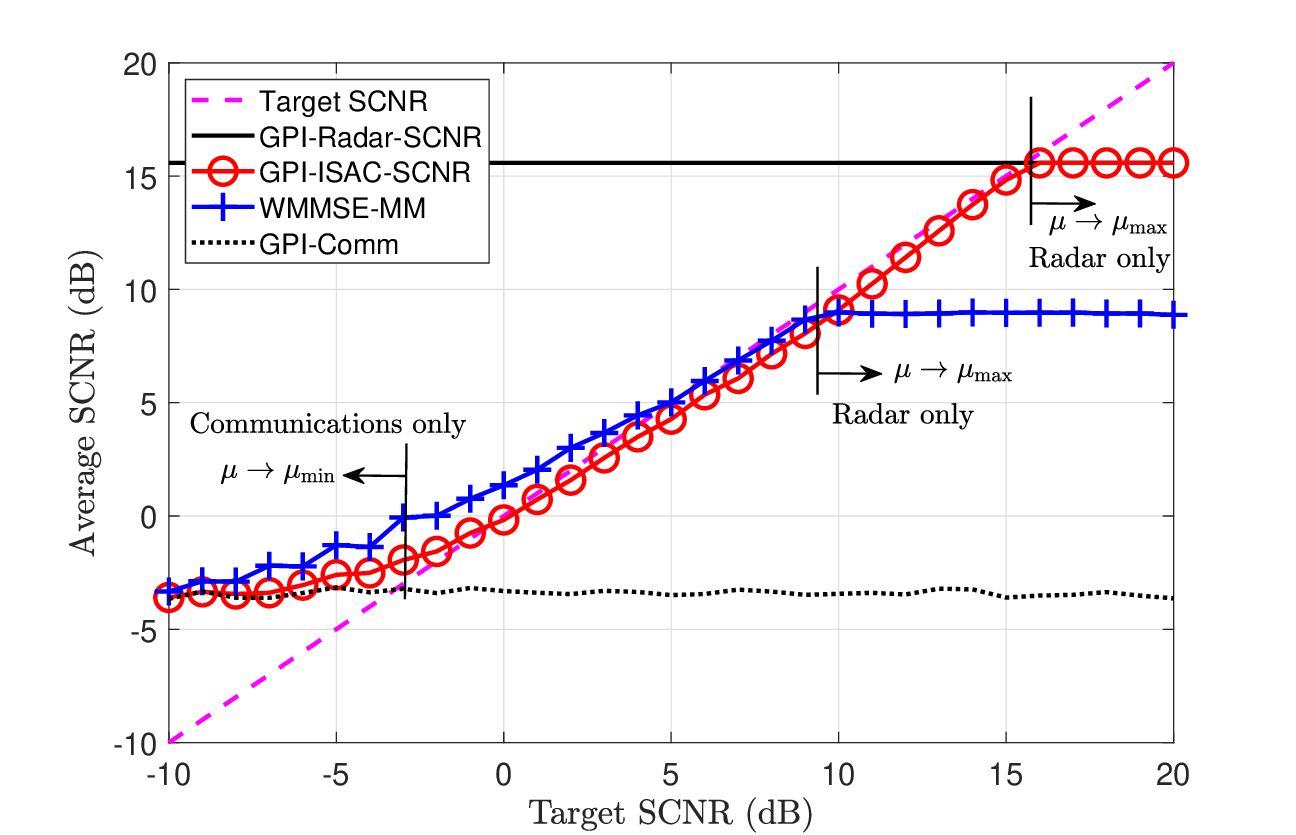}
    }
    \caption{(a) The ergodic sum SE and (b) the average SCNR with respect to the SNR for $N = 8$, $K = 4$, $M = 8$, and $25$ dB SNR.}
    \label{fig:VsScnr_N8K4SNR25_cl}
\end{figure}

Fig.~\ref{fig:VsMse_MSE} presents (a) the ergodic sum SE and (b) the average NMSE with respect to $T_{\rm nmse}$  for $N = 8$, $K = 4$, $M = 8$, and $40$ dB SNR.
It is shown in Fig.~\ref{fig:VsMse_MSE}(a) that the sum SE increases as the target NMSE increases for GPI-ISAC-MSE and SDR-based algorithms.
Beyond $-9.5$ dB, the SEs of GPI-ISAC-MSE and SDR-Optimal converge to that of GPI-Comm, which indicates that the NMSE target is too loose so that it is attainable without considering the constraint, i.e., $\mu = 0$.
When the target MSE is too stringent, the algorithms show a huge drop of the sum SE as they use the resource primarily to minimize the MSE. 
As shown in Fig.~\ref{fig:VsMse_MSE}(b), the drop occurs when the target MSE becomes infeasible ($< -11$~dB).
It is noticeable that the sum SE of GPI-ISAC-MSE is larger than that of SDR-Rank1 for the feasible target MSE.
The sum SE gap between GPI-ISAC-MSE and SDR-Rank1 becomes larger with the higher target MSE because the SDR-based algorithms present unnecessarily lower MSE than the target as shown in Fig.~\ref{fig:VsMse_MSE}(b).
Unlike the SDR-based algorithms, GPI-ISAC-MSE provides the higher sum SE while closely meeting the target MSE.
Fig.~\ref{fig:VsMse_MSE} reveals that the radar-oriented method is not preferable to use in terms of communications.

\subsection{SCNR Radar Performance Metric}

Now, we evaluate GPI-ISAC with the SCNR constraint (GPI-ISAC-SCNR) for a single-target case in which the intensity of clutters is comparable to the reflected beam from the target. 
To this end, we set reflection coefficients of the target and clutters as $|\beta^{\rm tar}| = |\beta_{j}^{\rm cl}| = 1$.
The target angle is $\theta^{\rm tar} = \pi/6$ radian and clutter angles are $\theta^{\rm cl}_j \in \{-\pi/3, -\pi/8, \pi/3\}$.
We consider $N = 8$, $K = 4$,  $M = 8$, and $25$ dB SNR.


\begin{figure} 
    \centering 
    {\includegraphics[width=0.8\columnwidth]{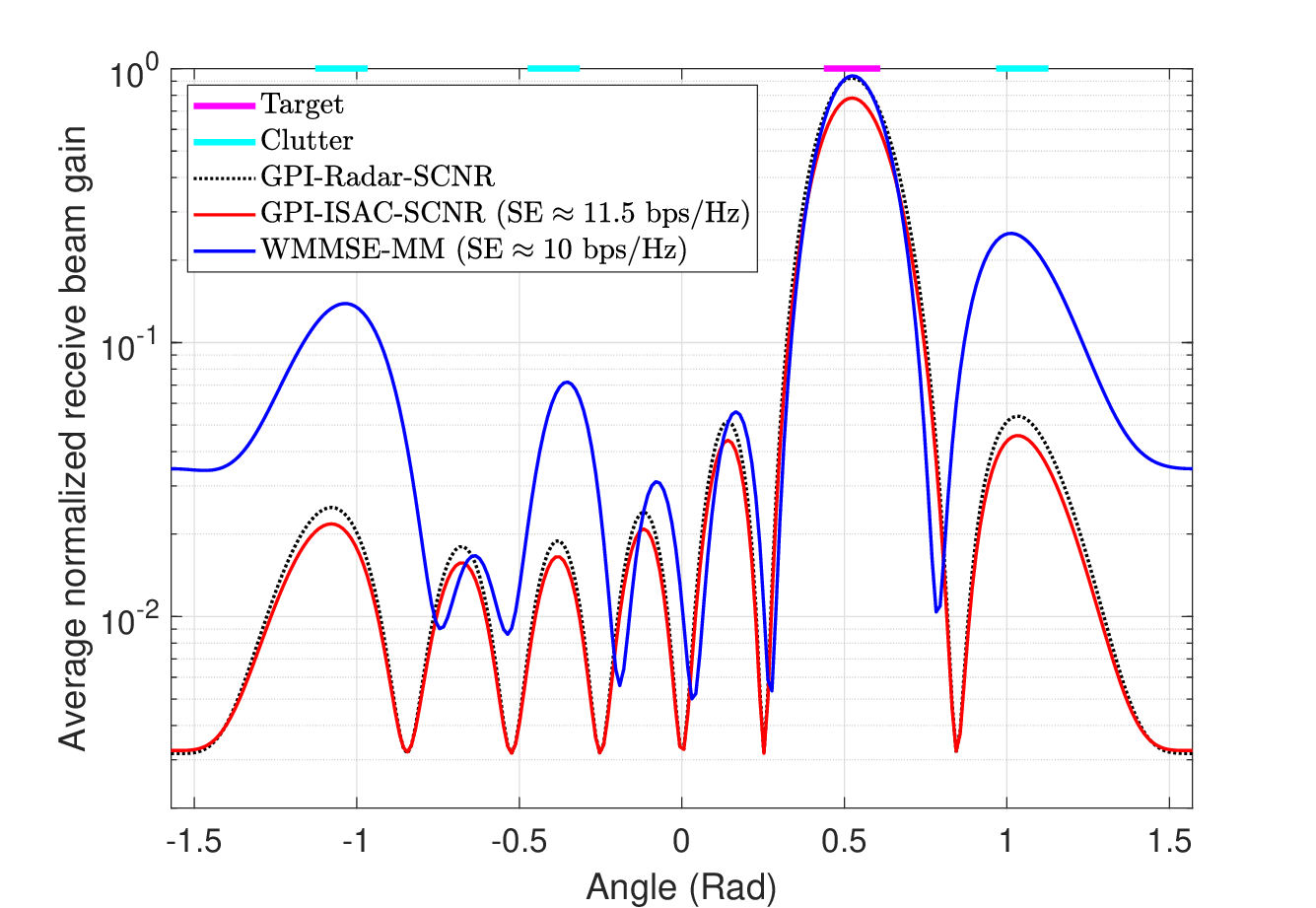}}
    \caption{Normalized receive beam patterns of GPI-based algorithms and WMMSE-MM  with the achieved sum SEs of $\sim\! 11.5\ {\rm bps/Hz}$  and $\sim\! 10\ {\rm bps/Hz}$ for GPI-ISAC-SCNR and WMMSE-MM, respectively.}
    \label{fig:RxBeam}
\end{figure}

We present the ergodic sum SE and the achieved SCNR performance for $25$ dB SNR in Fig.~\ref{fig:VsScnr_N8K4SNR25_cl}.
As shown in Fig.~\ref{fig:VsScnr_N8K4SNR25_cl}, GPI-ISAC-SCNR achieves a much higher sum SE than WMMSE-MM and also the linear precoders for the feasible target SCNR range.
When the target SCNR is low, the sum SEs of GPI-ISAC-SCNR and WMMSE-MM converge to different sum SEs.
In particular, the sum SE of GPI-ISAC-SCNR converge to that of GPI-Comm as the target SCNR decreases.
We notice that the convergence happens when the target SCNR becomes lower than the achieved SCNR of GPI-Comm, i.e., the target SCNR can be met with $\mu =0$.
Contrary to this, both the algorithms reveals rapid drop of the sum SE beyond a certain target SCNR because they put most of the resources to maximize the SCNR, i.e., $\mu \to \mu_{\rm max}$.
Such a significant drop of the sum SE is observed when the algorithms cannot satisfy the target SCNR ($> 16$ dB) as shown in Fig.~\ref{fig:VsScnr_N8K4SNR25_cl}(b).
With clutters, the feasible target SCNR range is wider for GPI-ISAC-SCNR than WMMSE-MM, which comes from the fact that WMMSE-MM does not take into account the clutters.
Therefore, GPI-ISAC-SCNR is more desirable to use than WMMSE-MM in the presence of non-negligible clutters.

Fig.~\ref{fig:RxBeam} illustrates a normalized receive beam gain over the angular grid: the normalized receive beam gain at each angle $\theta_\ell$ is computed as  
\begin{align}
    \frac{1}{P}\mathbb{E}[|{\bf a}_{\rm r}^{\sf H}(\theta_\ell){\bf y}_{\rm R}|^2] = {\bf a}_{\rm r}^{\sf H}\left({\bf G}{\bf F}{\bf F}^{\sf H}{\bf G}^{\sf H} + \frac{\sigma^2_{\rm R}}{P}{\bf I}_{N}\right){\bf a}_{\rm r},
\end{align}
where ${\bf G} = {\bf G}_{\rm tar}+{\bf G}_{\rm cl}$.
The achieved sum SEs are $\sim\! 11.5\ {\rm bps/Hz}$  and $\sim\! 10\ {\rm bps/Hz}$ for GPI-ISAC-SCNR and WMMSE-MM, respectively.
Even with the higher sum SE, GPI-ISAC-SCNR shows the better receive beam pattern than  WMMSE-MM, providing a comparable main lobe for the target angle with much lower side lobes; the achieved SCNR of GPI-ISAC-SCNR is approximately $15$ dB and that of WMMSE-MM is approximately $6$ dB.
Thus, the proposed method outperforms WMMSE-MM in terms of both beam design accuracy and SE.

    \begin{table}[!t]
    \begin{center}
        \caption{Normalized Computation Time by GPI-ISAC-MSE}
        \label{tab:table1_R3}
        \begin{tabular}{cccccc}
        \hline 
        Algorithms & GPI-ISAC-MSE & GPI-ISAC-SCNR & SDR & WMMSE  \\\hline
        Time & 1 &  0.15 &  8.73 &  1.06  \\ \hline
        \end{tabular}
    \label{tab:time}
    \end{center}
    
    \end{table}

Table~\ref{tab:time} shows the computation time normalized by that of the proposed MSE-based method for the considered algorithms for $N=8$, $K = 6$, $M=2$, $L = 128$, SNR $= 20$ dB, $T_{\rm nmse}=-16$ dB, and $T_{\rm scnr}=8$ dB.
    In the simulation, a single target at $0$ radian and a single clutter at $\pi/3$ radian are considered.
    We observe that the proposed method significantly reduces the computation time compared to the SDR-based approach (SDR-Rank1), while achieving the higher SE and better beam pattern as shown throughout the simulations.
    Although the WMMSE-MM algorithm's computation time is comparable to that of the proposed algorithm (GPI-ISAC-MSE), its performance is much lower than the proposed algorithms as observed throughout the simulations, and it cannot incorporate multiple targets or cluttering cases.
    This implies that, the proposed ISAC algorithms are more likely to better support real-time implementation with improved performance than the benchmarks.

\begin{figure} 
    \centering 
    \subfigure[No cluttering (single-target \& multi-target)]{\includegraphics[width=1.\columnwidth]{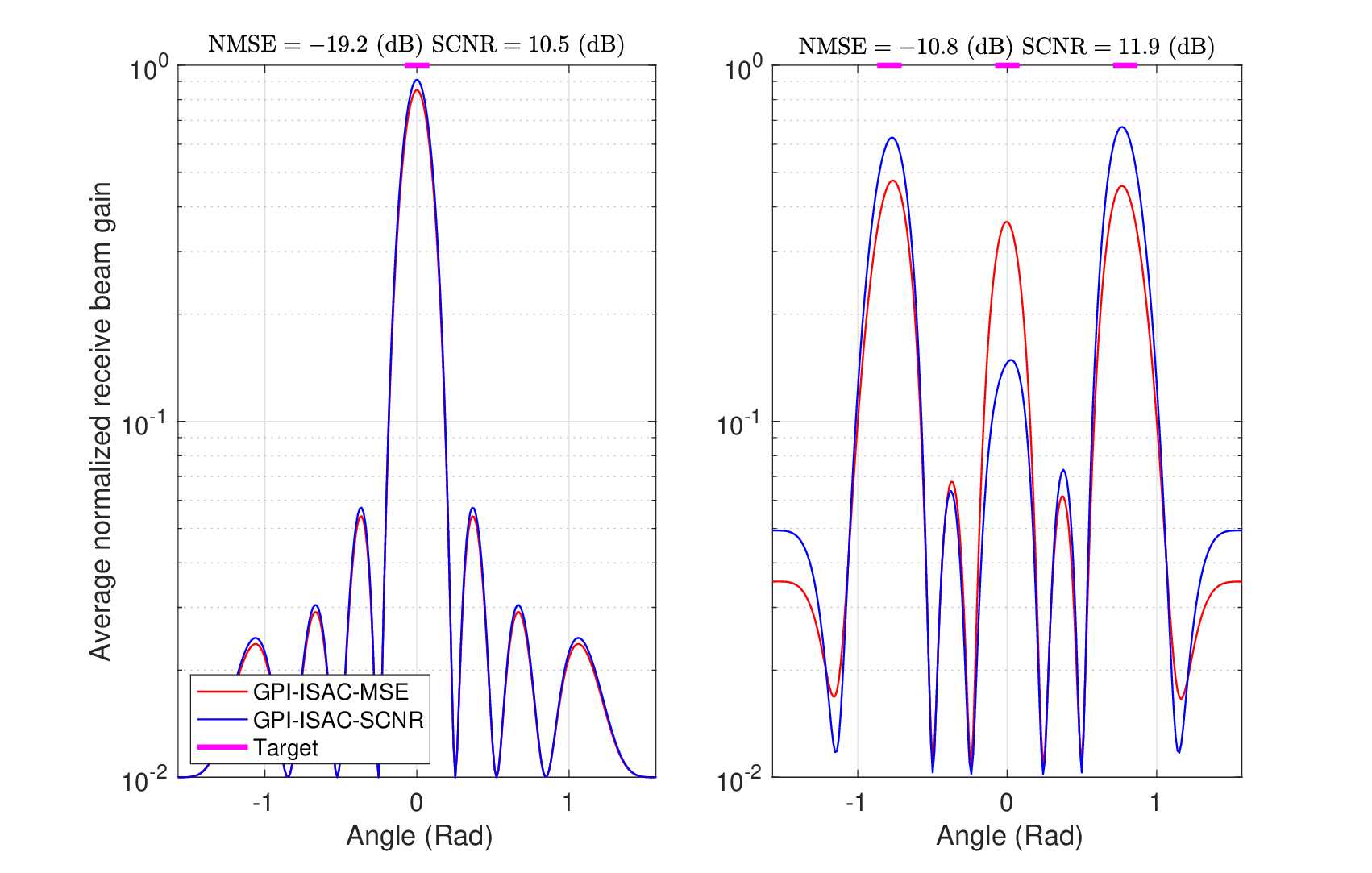}}
    \\
    \centering 
    \subfigure[Cluttering (single-target \& multi-target)]{\includegraphics[width=1.\columnwidth]{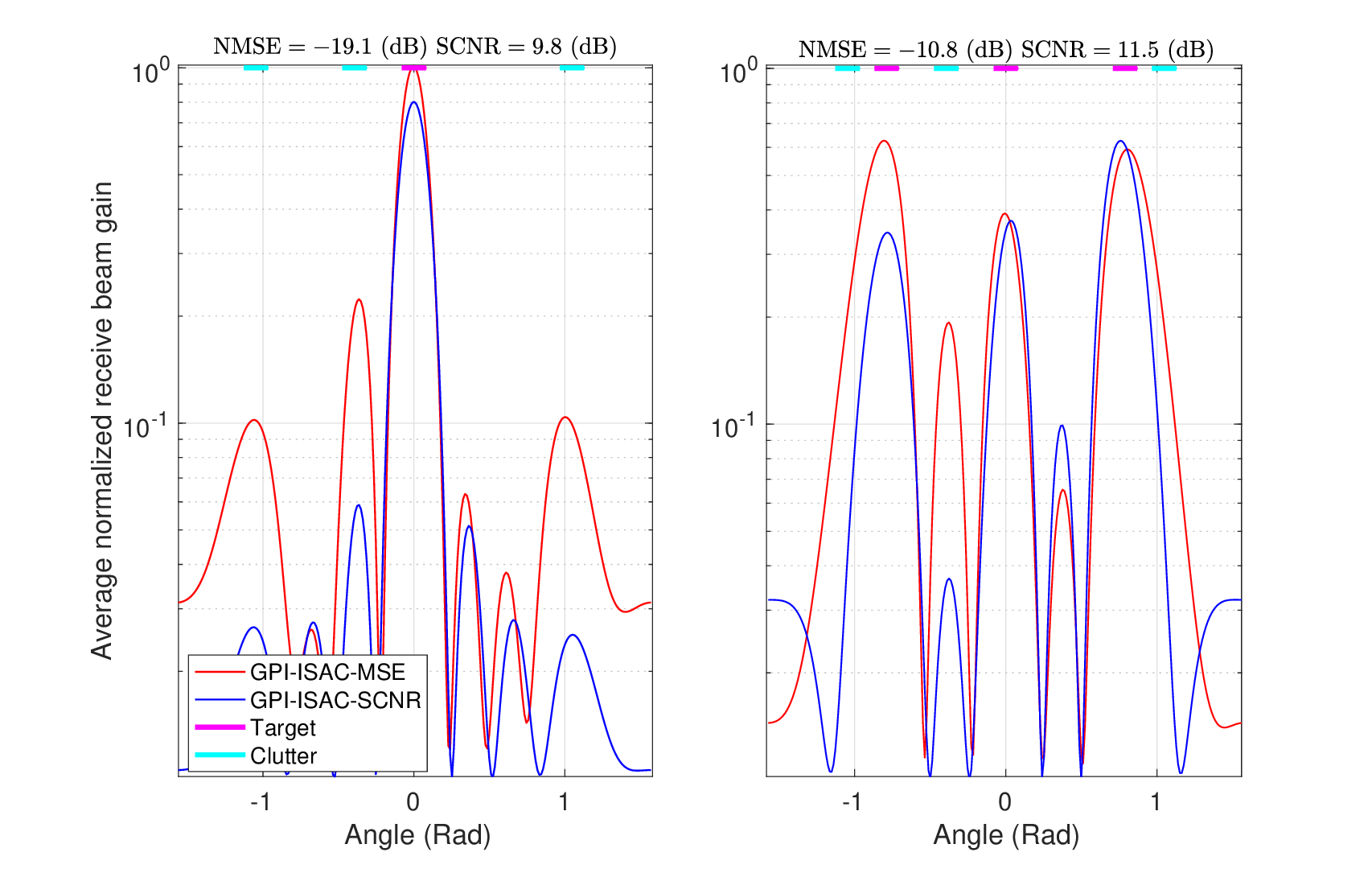}}
    \caption{Radar performance with regard to the number of targets and the intensity of cluttering at the achieved sum SE of  $\sim \!\! 10\ {\rm bps/Hz}$. The achieved NMSE of GPI-ISAC-MSE and SCNR of GPI-ISAC-SCNR are also presented.}    
    \label{fig:comparison}
\end{figure}

\subsection{Comparison between MSE and SCNR metrics}

Finally, we analyze the effect of using the MSE and SCNR as the radar performance metric for single-target, multi-target, no cluttering, and cluttering cases.
In Fig.~\ref{fig:comparison}, the receive beam gains of GPI-ISAC-MSE and GPI-ISAC-SCNR are plotted for $N=8$, $K=6$, $M=8$, and $20$ dB SNR at the achieved sum SE of $\sim \!\! 10\ {\rm bps/Hz}$.
When there is no or negligible clutter, GPI-ISAC-MSE and GPI-ISAC-SCNR provide almost similar beam patterns for the single-target case, whereas GPI-ISAC-MSE presents a better pattern with better-balanced main lobes and lower side lobes than GPI-ISAC-SCNR for the multi-target case as shown in Fig.~\ref{fig:comparison}(a).
Such a result corresponds to the intuition that the MSE radar metric is adequate to design a multi-target beam without clutters.

In Fig.~\ref{fig:comparison}(b), the receive beam patterns are illustrated for the non-negligible cluttering environment.
With the presence of a single target, GPI-ISAC-SCNR outperforms GPI-ISAC-MSE as its beam pattern shows similar main lobe with much lower side lobes than that of GPI-ISAC-MSE.
This also aligns with the intuition that the SCNR metric is suitable for a single-target case considering non-negligible cluttering.
A multi-target case, however, the beam patterns of both GPI-ISAC-MSE and GPI-ISAC-SCNR reveal trade-off:  GPI-ISAC-MSE presents more even or stronger main lobes with higher side lobe whereas  GPI-ISAC-SCNR shows lower side lobe  with more uneven or weaker main lobes  compared to each other.
Accordingly, for such a harsh environment, 
either metric can be used depending on the types of radar systems. 

We further provide a quantitative analysis on the radar metrics: MSE and SCNR by evaluating the detection probability of the proposed radar-only algorithms.
The proposed algorithm for the communication-only case is also evaluated for comparison.
For the evaluation, we consider $N = 8$, $K = 4$, and $M = 8$. 
We set reflection coefficients of the targets and clutters as $|\beta^{\sf tar}_i| = 1$ and $ |\beta_j^{\sf cl}| = 10$, $\forall i,j$ to consider strong cluttering scenarios for comparison of MSE and SCNR metrics. 
The detection probability for the case of nonfluctuating target in \cite{de2008code}  is adopted with the similar setup as in \cite{de2008code}. 
In all cases, we consider the target probability of false alarm to be $10^{-4}$.
For multi-target cases, we present the worst detection probability
and the average detection probability.

\begin{figure} 
    \centering 
    \subfigure[No cluttering]{\includegraphics[width=0.85\columnwidth]{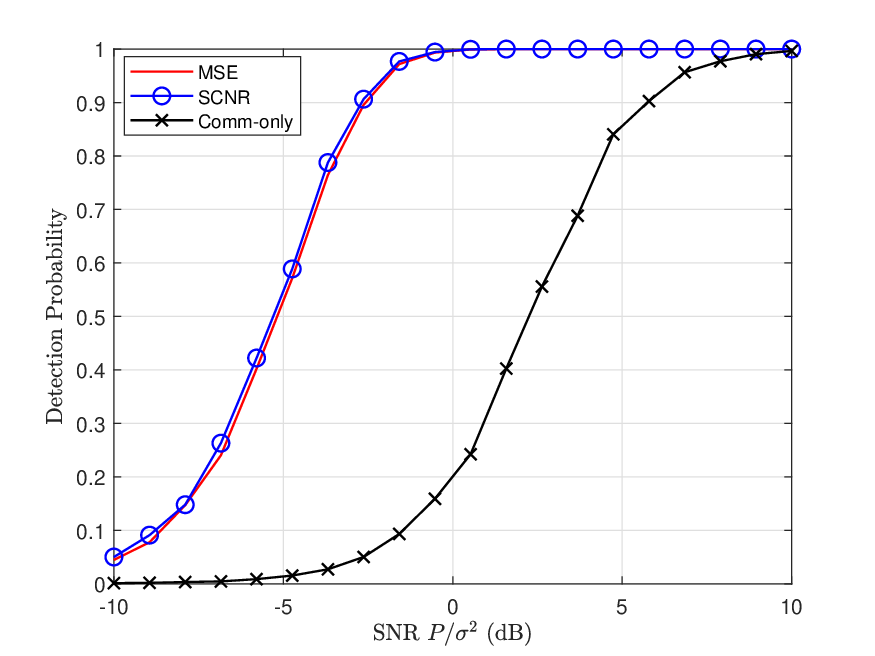}}\\
    \centering 
    \subfigure[Cluttering]{\includegraphics[width=0.85\columnwidth]{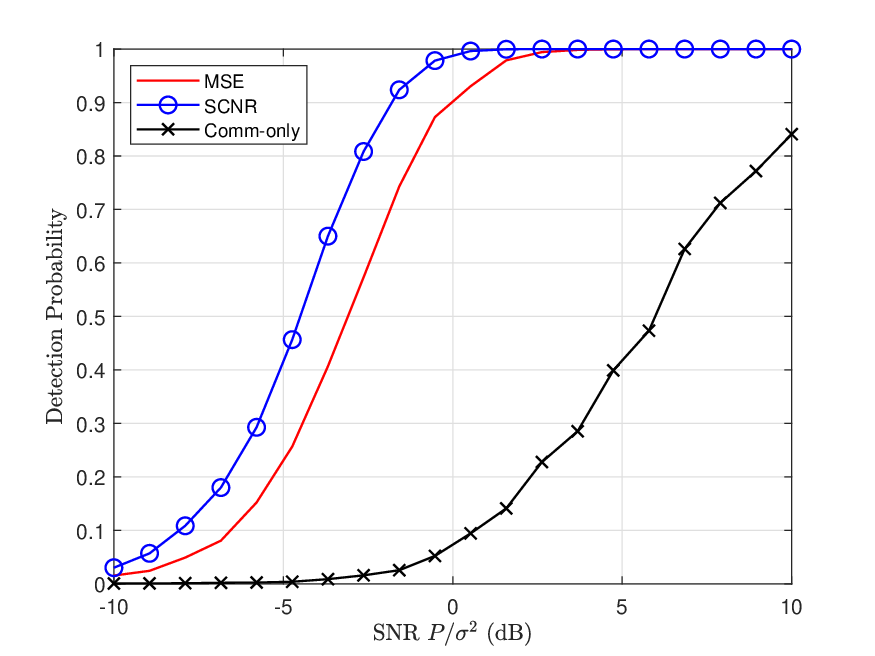}}
    \caption{The probability of detection for single-target cases with $N = 8$, $K = 4$, and $M = 8$.}
    \label{fig:Pdetect}
\end{figure}

Fig.~\ref{fig:Pdetect} shows the detection probability for the single-target cases with respect to the transmit SNR.
The considered target angle is $0$ radian and cluttering angles are $(-\pi/8, \pi/3)$  radian.
As expected, the MSE and SCNR-based algorithms achieve similar detection probability for the single-target only scenario which is shown in Fig.~\ref{fig:Pdetect}(a). 
Fig.~\ref{fig:Pdetect}(b), however, reveals that the SCNR-based algorithm provides the higher detection probability than the MSE-based algorithm for the single target with strong cluttering scenario because it takes the cluttering into account when optimizing the beamforming.
The observations from Fig.~\ref{fig:Pdetect} exactly align with the analysis results from Fig.~\ref{fig:comparison}.

\begin{figure} 
    \centering 
    \subfigure[No cluttering]{\includegraphics[width=0.85\columnwidth]{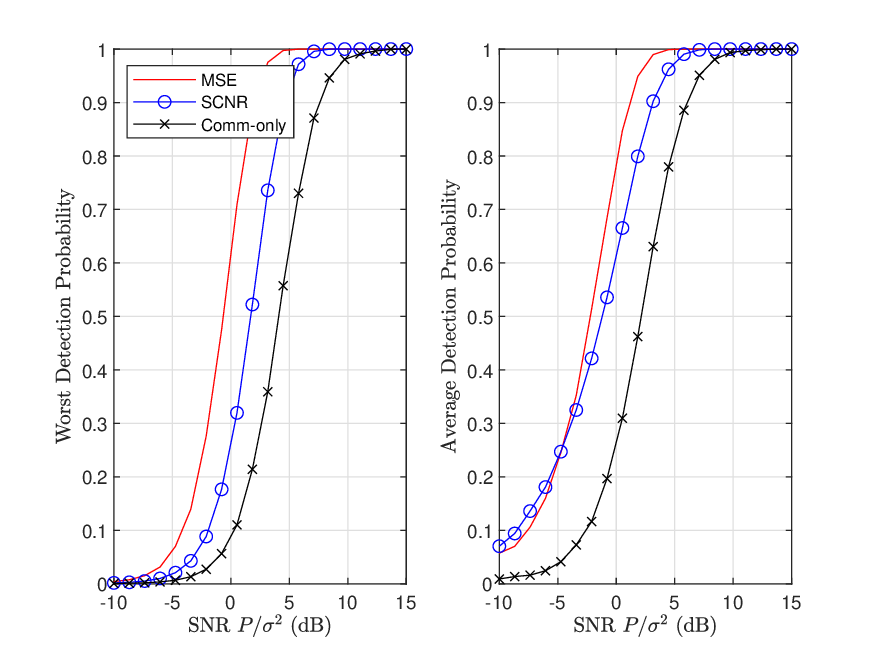}}\\
    \centering 
    \subfigure[Cluttering]{\includegraphics[width=0.85\columnwidth]{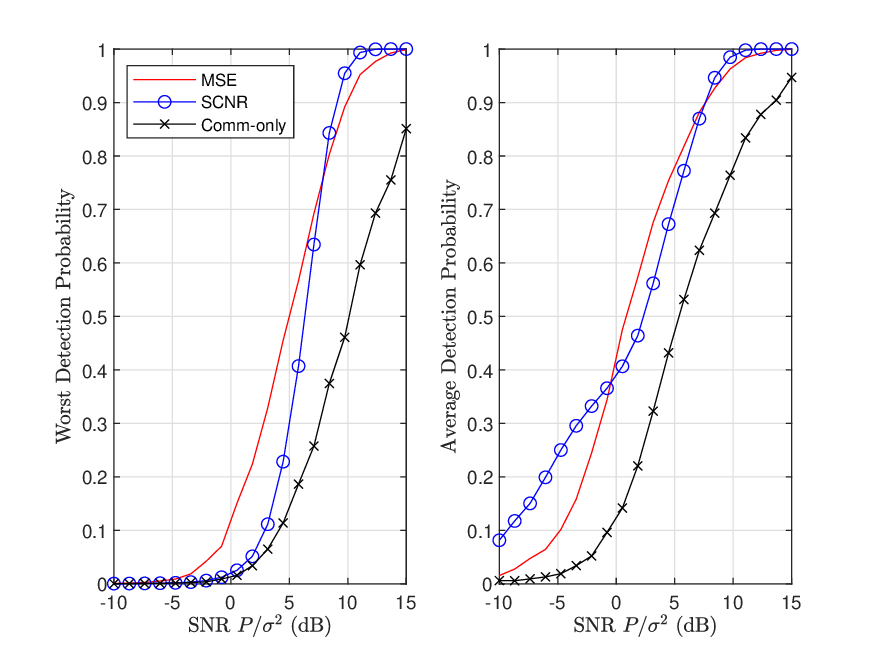}}
    \caption{The probability of detection for three-target cases with $N = 8$, $K = 4$, and $M = 8$.}
    \label{fig:Pdetect_multi}
\end{figure}
Now, let us consider multi-target scenarios.
We consider target angles as $(-\pi/3,0,\pi/4)$  radian.
For cluttering angles, we consider $(-\pi/8, \pi/3)$  radian.
In Fig.~\ref{fig:Pdetect_multi}(a), the multi-target beamforming case, the MSE-based algorithm attains noticeable improvement in both the worst and average detection probabilities compared to the SCNR-based algorithm; unlike the MSE-based algorithm, the SCNR-based algorithm cannot guarantee the uniform beam gains for all targets, which is observed in Fig.~\ref{fig:comparison}.
When strong cluttering is considered as in Fig.~\ref{fig:Pdetect_multi}(b),  the MSE-based algorithm also presents the better worst detection probability for the low-to-medium SNR regime.
Regarding the average detection probability, however, the SCNR-based algorithm achieves better performance in the low SNR regime.
Such an observation from Fig.~\ref{fig:Pdetect_multi} indicates that thanks to the relatively more uniform beam gain for all targets, the MSE-based algorithm balances the detection probability of all target whereas the SCNR-based algorithm can maximize the detection probability of the subset of the targets without caring the worst detection probability.
In this regards, when there are multiple targets with strong cluttering, both the MSE-based algorithm and SCNR-based algorithms can be selectively used depending on the sensing scenario and requirements.
The observations from Fig.~\ref{fig:Pdetect_multi} also  match with the  analysis results from Fig.~\ref{fig:comparison}.
The analysis results are summarized in Table~\ref{tab:table1}.

\begin{table}[!t]
{\color{black}{
  \begin{center}
    \caption{A Preferred Radar Performance Metric for Different Cases}\vspace{-0.5em}
    \label{tab:table1}
    \begin{tabular}{c|c|c} 
    \hlineB{2}
    {\textbf{Case}} & {\textbf{Low Clutter Intensity}} & {\textbf{High Clutter Intensity} } \\
    \hline
      {\textbf{Few Targets}} & {\text{MSE, SCNR}} & {\text{SCNR}} \\
    \hline
      {\textbf{Many Targets}} & {\text{MSE}}  &  {\text{MSE, SCNR}}   \\
    \hlineB{2}
    \end{tabular}
  \end{center}
  }}\vspace{-1.5em}
\end{table}

\subsection{Robustness to CSIT Imperfection}
\begin{figure} 
 \centering 
   {\includegraphics[width=0.85\columnwidth]{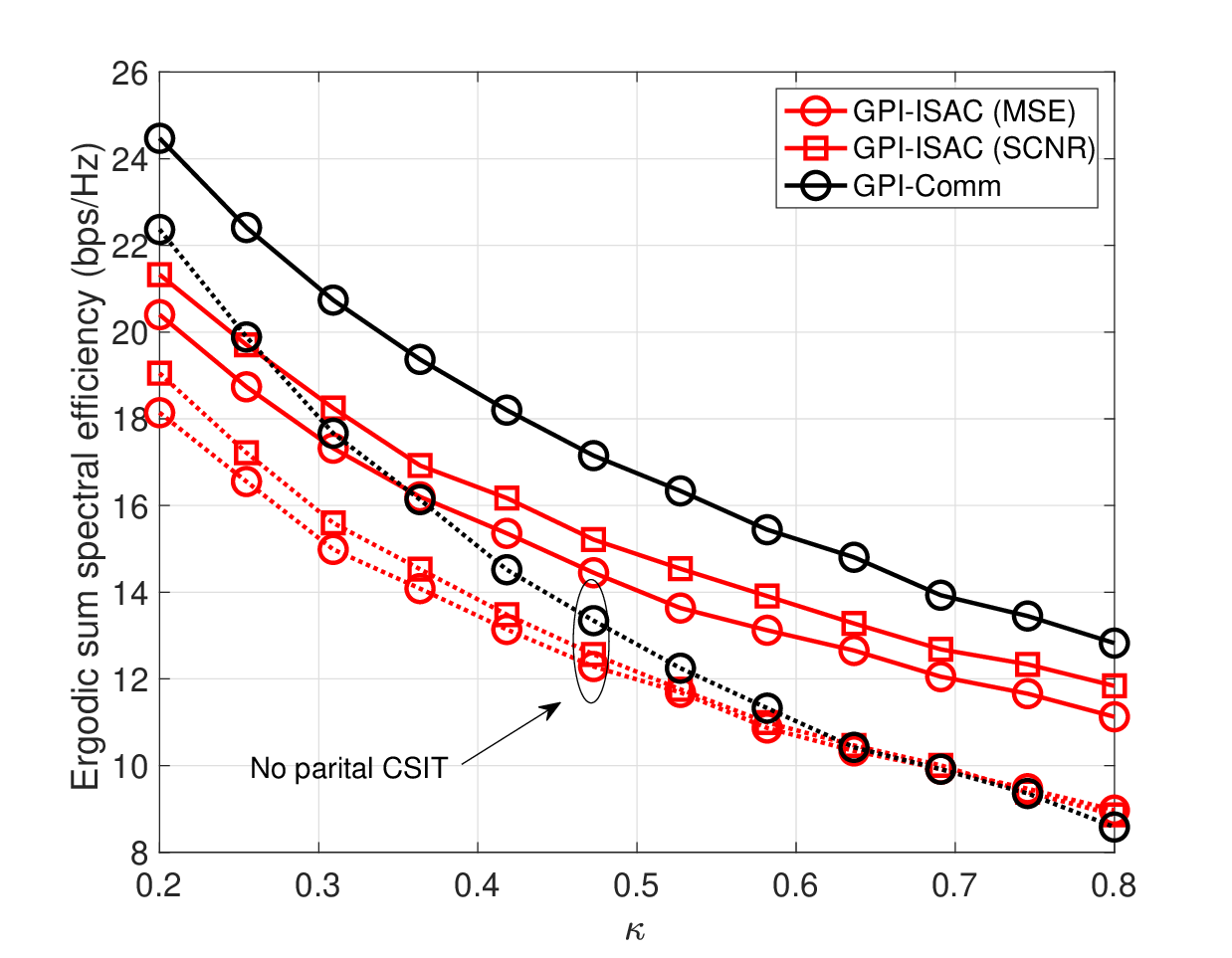}}
    \caption{The achieved spectral efficiency for $N = 8$, $K = 4$, $M = 8$,  ${\rm SNR} = 25$ dB, $T_{\rm nmse} = -12$ dB, and $T_{\rm scnr} = 14$ dB.}
    \label{fig:ChEstError}
\end{figure}

The robustness of the proposed ISAC beamforming algorithms to the imperfect CSIT is evaluated in Fig.~\ref{fig:ChEstError}  for $N = 8$, $K = 4$, $M = 8$,  ${\rm SNR} = 25$ dB, $T_{\rm nmse} = -12$ dB, and $T_{\rm scnr} = 14$ dB.
We consider two cases: the proposed algorithms with utilizing the channel error covariance matrix ${\pmb \Phi}_k$ and without utilizing it, i.e., setting ${\pmb \Phi}_k =  {\bf 0}$ in the algorithm.
As shown in Fig.~\ref{fig:ChEstError}, the proposed algorithms with the error matrix utilization achieve higher SEs than those without the error matrix. 
This observation demonstrates that the proposed algorithms are more robust to the channel estimation error by properly leveraging the partial channel information. 
In particular, the communication-only case shows greater improvement when using the error information, 
as this case solely focuses on maximizing the SE.

\section{Conclusion}

In this paper, we proposed a joint beamforming framework for integrated sensing and communications systems with imperfect CSIT.
Using a MSE or SCNR  metric as a radar constraint, we formulated a communication-oriented problem to maximize a sum SE while satisfying the  MSE or SCNR constraint.
To solve the problem,
we first reformulated the problem in terms of a vectorized joint beamformers, and then identified the KKT stationarity condition.
Interpreting the condition as a functional eigenvalue problem, a GPI method was employed to derive the best local optimal solution.
In simulations, the proposed GPI-ISAC algorithms demonstrated the superior communications and radar performance over the state-of-the-art ISAC methods with the imperfect CSIT by achieving lower computational time.
The compatibility of the different radar metrics was also analyzed, showing that the MSE metric is effective in multi-target and no-cluttering environments and the SCNR metric is useful in single-target and intense-cluttering environments. 
Therefore, it is concluded that the proposed ISAC beamforming framework is  comprehensive, robust, and low-complexity with high performance.

\appendices

\section{Proof of Lemma \ref{lem:main}}
\label{append:lem1}

We first normalize the MSE constraint with respect to the transmit power $P$ to make the scale of the constraint comparable to the objective function as 
\begin{align}
    \label{eq:MSE_norm}
    \frac{1}{LP^2} \sum_{\ell = 1}^{L} |  \bar {\bf{f}}^{\sf H} {\bf{D}}(\theta_\ell) \bar {\bf{f}} - \bar {\bf{f}}^{\sf H} {\bf{A}}(\theta_\ell) \bar {\bf{f}} |^2 \leq \frac{T_{\rm mse}}{P^2}.
\end{align}
The Lagrangian function of \eqref{eq:prob_reformulation_omit_power} with the MSE in \eqref{eq:MSE_norm} is 
\begin{align} 
    \nonumber
    L(\bar {\bf{f}}; \mu) &= \log_2 \prod_{k = 1}^{K} \left(\frac{\bar{\bf{f}}^{\sf H}{\bf{B}}_k \bar {\bf{f}}}{\bar{\bf{f}}^{\sf H} {\bf{C}}_k\bar{\bf{f}}} \right) +
    \\\label{eq:lagrangian}
    &\quad  \mu \left(\frac{T_{\rm mse}}{P^2} - \frac{1}{LP^2} \sum_{\ell = 1}^{L} |  \bar {\bf{f}}^{\sf H} {\bf{D}}(\theta_\ell) \bar {\bf{f}} - \bar {\bf{f}}^{\sf H} {\bf{A}}(\theta_\ell) \bar {\bf{f}} |^2  \right)
    \\
    & = \log_2\lambda(\bar {\bf{f}}),
\end{align}
where $\mu$ is the Lagrangian multiplier and $\lambda(\bar {\bf{f}})$ is defined in \eqref{eq:lambda}.  
The stationarity condition requires 
\begin{align} 
    \label{eq:derivative}
         \frac{\partial L(\bar {\bf{f}}, \mu)}{\partial \bar {\bf{f}}}  =  \frac{\partial L(\bar {\bf{f}}, \mu)}{\partial \lambda(\bar {\bf{f}})}\frac{\partial \lambda(\bar {\bf{f}})}{\partial \bar {\bf{f}}}  =\frac{1}{ \lambda(\bar {\bf{f}})\log2} \frac{\partial \lambda(\bar {\bf{f}})}{\partial \bar {\bf{f}}} =0.
\end{align}
Accordingly, we find the condition for $\frac{\partial \lambda(\bar {\bf{f}})}{\partial \bar {\bf{f}}}= 0$ which is equivalent to $\eqref{eq:derivative}$.
With direct matrix calculus and simplification, we have
\begin{align}
    \nonumber
    \frac{\partial \lambda(\bar {\bf{f}})}{\partial \bar {\bf{f}}}= & 2\lambda(\bar {\bf{f}})\left[\sum_{k = 1}^{K}  \Bigg(\frac{{\bf{B}}_{k} \bar {\bf{f}} }{\bar {\bf{f}}^{\sf H} {\bf{B}}_{k} \bar {\bf{f}}} - \frac{{\bf{C}}_{k} \bar {\bf{f}}}{\bar {\bf{f}}^{\sf H} {\bf{C}}_{k} \bar {\bf{f}}} \Bigg) \right.
    \\\nonumber
    & \left.+   \frac{2\mu \log2}{LP^2} \sum_{\ell = 1}^{L}\Bigg( \left(\bar {\bf{f}}^{\sf H} {\bf{A}}(\theta_\ell) \bar {\bf{f}} -  \bar {\bf{f}}^{\sf H} {\bf{D}}(\theta_\ell) \bar {\bf{f}} \right)  {\bf{D}}(\theta_\ell) \bar {\bf{f}}  \right.
    \\ \label{eq:lambda_derive}
    & \left.-  \left(\bar {\bf{f}}^{\sf H} {\bf{A}}(\theta_\ell) \bar {\bf{f}} -  \bar {\bf{f}}^{\sf H} {\bf{D}}(\theta_\ell) \bar {\bf{f}} \right) {\bf{A}}(\theta_\ell) \bar {\bf{f}}\Bigg)\right] = 0. 
\end{align}
Rearranging the condition in \eqref{eq:lambda_derive} carefully, we have
\begin{align}
    \nonumber
    & \lambda_{\sf num}(\bar {\bf f})\left[ \sum_{k = 1}^{K}    \frac{{\bf{B}}_{k}  }{\bar {\bf{f}}^{\sf H} {\bf{B}}_{k} \bar {\bf{f}}}+ \right.
    \\\nonumber
    &\left. \frac{2 \mu\log2}{LP^2} \sum_{\ell = 1}^{L} \left\{ ( \bar {\bf{f}}^{\sf H} {\bf{D}}(\theta_\ell) \bar {\bf{f}}) {\bf{A}}(\theta_\ell) + (\bar {\bf{f}}^{\sf H} {\bf{A}}(\theta_\ell) \bar {\bf{f}})  {\bf{D}}(\theta_\ell) \right\}    \right]  \bar {\bf{f}} 
    \\ \nonumber
    & = \lambda_{\sf den}(\bar {\bf f})\left[ \sum_{k = 1}^{K}    \frac{{\bf{C}}_{k}  }{\bar {\bf{f}}^{\sf H} {\bf{C}}_{k} \bar {\bf{f}}} + \right.
    \\ \nonumber 
    &\left.  \frac{2 \mu \log2}{LP^2} \sum_{\ell = 1}^{L} \left\{ ( \bar {\bf{f}}^{\sf H} {\bf{D}}(\theta_\ell) \bar {\bf{f}}) \alpha {\bf{D}}(\theta_\ell) + (\bar {\bf{f}}^{\sf H} {\bf{A}}(\theta_\ell) \bar {\bf{f}}) {\bf{A}}(\theta_\ell) \right\}    \right] \lambda({\bar {\bf f}}) \bar {\bf{f}}.
\end{align}
This completes the proof. 
\hfill $\blacksquare$

\section{Proof of Lemma \ref{lem:main2}}
\label{append:lem2}

We first convert the SCNR constraint into a log-scale to make the scale of the constraint comparable to the sum SE as 
\begin{align}
    \label{eq:SCNR_log}
     \log_2 \left(\frac{\bar {\bf f}^{\sf H} {\bar {\bf G}}_{\rm tar}\bar {\bf f}}{\bar {\bf f}^{\sf H}{\bar {\bf G}}_{\rm cl}\bar {\bf f}}\right) \ge \log_2 T_{\rm scnr}.
\end{align}
The Lagrangian function of \eqref{eq:prob_reformulation_scnr} with \eqref{eq:SCNR_log} is 
\begin{align} 
    \nonumber
    \tilde L(\bar {\bf{f}}; \mu) &= \log_2 \prod_{k = 1}^{K} \left(\frac{\bar{\bf{f}}^{\sf H}{\bf{B}}_k \bar {\bf{f}}}{\bar{\bf{f}}^{\sf H} {\bf{C}}_k\bar{\bf{f}}} \right) + \mu \left(  \log_2 \left(\frac{\bar {\bf f}^{\sf H} {\bar {\bf G}}_{\rm tar}\bar {\bf f}}{\bar {\bf f}^{\sf H}{\bar {\bf G}}_{\rm cl}\bar {\bf f}}\right) - \log_2 T_{\rm scnr}\right)
    \\\label{eq:lagrangian_scnr}
    & = \log_2\eta(\bar {\bf{f}}),
\end{align}
where $\mu$ is the Lagrangian multiplier and $\eta(\bar {\bf{f}})$ is defined in \eqref{eq:eta}.  
The stationarity condition requires 
\begin{align} 
    \label{eq:derivative_scnr}  
    \frac{\partial \tilde L(\bar {\bf{f}}, \mu)}{\partial \bar {\bf{f}}}  =  \frac{\partial \tilde L(\bar {\bf{f}}, \mu)}{\partial \eta(\bar {\bf{f}})}\frac{\partial \eta(\bar {\bf{f}})}{\partial \bar {\bf{f}}}  = \frac{1}{ \eta(\bar {\bf{f}})\log2} \frac{\partial \eta(\bar {\bf{f}})}{\partial \bar {\bf{f}}}.
\end{align}
Accordingly, we find the condition for $\frac{\partial \eta(\bar {\bf{f}})}{\partial \bar {\bf{f}}}= 0$.
\begin{align}
    \nonumber
    \frac{\partial \eta(\bar {\bf{f}})}{\partial \bar {\bf{f}}}&\!=\! 2\eta(\bar {\bf{f}})\left[\sum_{k = 1}^{K}  \Bigg(\frac{{\bf{B}}_{k} \bar {\bf{f}} }{\bar {\bf{f}}^{\sf H} {\bf{B}}_{k} \bar {\bf{f}}} \!-\! \frac{{\bf{C}}_{k} \bar {\bf{f}}}{\bar {\bf{f}}^{\sf H} {\bf{C}}_{k} \bar {\bf{f}}} \Bigg) \!+ \! \mu \Bigg(\frac{ {\bar {\bf G}}_{\rm tar}\bar {\bf f} }{\bar {\bf f}^{\sf H}{\bar {\bf G}}_{\rm tar}\bar {\bf{f}} } \!-\! \frac{ {\bar {\bf G}}_{\rm cl}\bar {\bf f} }{\bar {\bf f}^{\sf H}{\bar {\bf G}}_{\rm cl}\bar {\bf{f}} }\Bigg)\right] 
    \\ \label{eq:eta_derive}
    &= 0. 
\end{align}
Rearranging the condition in \eqref{eq:eta_derive} carefully, we have
\begin{align}
    &\eta_{\sf num}(\bar {\bf f})\left( \sum_{k = 1}^{K}    \frac{{\bf{B}}_{k}  }{\bar {\bf{f}}^{\sf H} {\bf{B}}_{k} \bar {\bf{f}}} + \mu \frac{ {\bar {\bf G}}_{\rm tar}}{\bar {\bf f}^{\sf H}{\bar {\bf G}}_{\rm tar}\bar {\bf{f}} }   \right)  \bar {\bf{f}}  
    \\
    &= \eta_{\sf den}(\bar {\bf f})\left( \sum_{k = 1}^{K}    \frac{{\bf{C}}_{k}  }{\bar {\bf{f}}^{\sf H} {\bf{C}}_{k} \bar {\bf{f}}} + \mu  \frac{ {\bar {\bf G}}_{\rm cl}}{\bar {\bf f}^{\sf H}{\bar {\bf G}}_{\rm cl}\bar {\bf{f}} } \right) \eta({\bar {\bf f}}) \bar {\bf{f}}.
\end{align}
This completes the proof. 
\hfill $\blacksquare$

\bibliographystyle{IEEEtran}
\bibliography{ref_MIMO_DFRC}

\end{document}